# Bioinspired nested-isotropic lattices with tuneable anisotropy for additive manufacturing

B. Ramalingaiah[1], B. Panda[1], S. Kumar[2,3*]

[1] Department of Mechanical Engineering; Indian Institute of Technology Guwahati, Guwahati-781039, Assam, India,

[2] James Watt School of Engineering, University of Glasgow, Glasgow, G12 8QQ, UK,

[3] Glasgow Computational Engineering Centre, University of Glasgow, Glasgow, G12 8LT, UK

*Email: msv.kumar@glasgow.ac.uk

## Abstract

This study presents innovative nested-isotropic lattices for additive manufacturing, drawing inspiration from bio-architectures found in cortical bone osteons, golden spirals, and fractals. These lattices provide tuneable anisotropy by integrating architectural elements like "nesting orders (NOs)" and corresponding "nesting orientations (NORs)," along with repetitive self-similar X-cross struts and three four-fold axes of symmetry, resulting in a wide spectrum of lattice designs. Nine mono-nest and twenty multi-nest lattices, along with 252 parametric variations, are realized. The relative density $\bar{\rho}$ and surface area density $\bar{S}$ are calculated. Employing finite element-based numerical homogenization, elastic stiffness tensors are estimated to evaluate the anisotropic measure - Zener ratio $Z$ and elastic modulus $\bar{E}$ for all lattice designs. The mono-nest lattices generated considering higher NOs and respective NORs exhibit a transition from shear dominant to tensile/compression dominant (TCD) anisotropic behavior and their strut size variations show a strong influence on $\bar{\rho}$, $\bar{S}$ and $\bar{E}$. In contrast, multi-nest lattices exhibit isotropic and neo-isotropic characteristics, with strut size mismatch exerting more influence on $Z$. Increasing NOs and NORs result in isotropic or TCD behavior for most multi-nest lattices, with strut size mismatch leading to many isotropic lattices. These bio-inspired nested lattices, coupled with advancements in additive manufacturing, hold potential for diverse applications.

## 1. Introduction

Over the last few decades, utilization of cellular materials and structures has increased in a wide array of applications including biomedical, aerospace, automobile, protective devices, heat exchangers, defence and thermal protection systems due to their exceptional performance attributes such as lightweight, high weight-specific stiffness, strength and energy absorption, impact resistance, acoustic and vibration damping, negative Poisson's ratio, high surface area density and heat dissipation[1–4]. Initially, foams were employed due to their near isotropic behavior, particularly under compression and shear but their usage was limited due to low structural efficiency (lower weight-specific strength and stiffness and energy absorption), lack of multifunctionality and difficulties in manufacturing[5–8]. This led to the development of lattice structure[9–17] - an ordered cellular structure comprising repetitive unit cell architectures. Emerging additive manufacturing techniques enable the fabrication of complex cellular geometries across length scales[18–29].

Most extant studies explored the design of lattice structures mimicking elementary crystal structures[30] such as simple cubic (SC)[31], face-centered cubic (FCC)[32] and body-centered cubic (BCC)[33] due to their simpler design and improved mechanical performances compared to foams. The proliferation of lattice geometries such as octet[34], diamond[35], rhombic dodecahedron[36] and octahedron[37] showed further improvements in mechanical properties and other attributes relevant to functional requirements. Another set of studies focused on the usage of implicit functions such as triply periodic minimal surfaces (TPMS) of gyroid, neovius, I-graph-wrapped package graph (IWP), fischer-koch-S and primitive sheet[38]. Even though these lattices exhibit superior performance, the usage of complex mathematical functions and difficulties in designing and fabricating them with higher porosities hinder their applications. While all the aforementioned architectures are utilized for suitable applications, they lack isotropy, a crucial characteristic for numerous engineering applications to withstand complex loadings and evenly distribute them in all directions[39]. Especially in load-bearing bone implants, the ideal lattice structures should possess: 1) isotropic behavior to distribute the stress evenly[39], 2) highly porous architecture to accommodate the blood vessels and nerves as well as enhance osseointegration[40], 3) high surface area density to improve cellular activities such as cell adhesion, growth, and proliferation[40], 4) low mismatch in stiffness with the native bone to avoid the stress shielding[40], and 5) high fatigue resistance[41,42]. Most of the conventional lattice structures lack these attributes. This disparity raises concerns among researchers, prompting us to explore lattice structures that encompass multiple performance attributes tailored to specific applications.

A few recent studies explored the design of isotropic lattice structures (ILSs), exploiting four common design strategies: a) combining two or more basic unit cells, such as SC, BCC, FCC, octet and octahedron, with complementary stiffness where SC and Octet unit cells are considered as reference and combined with unit cells that exhibit anisotropic behavior, b) adjustment of their respective ligament dimensions or geometries by considering the different ratios of sizes and shapes of struts, shells, and plates[39,43–45,38], c) symmetric operation of basic unit cells where the unit cells are tessellated in such a way that one can get ILSs[44] and d) topology optimization is yet another way of realizing ILSs[46], but the inherent complexity and manufacturability challenges associated with the optimized designs hinder their potential. Although, these methods brought a fruitful set of ILSs and control of anisotropic behavior but their performances and applications are limited by the usage of common conventional unit cells and design methods. To address this research gap, this study aims to develop novel architectures to generate ILSs with tuneable and controllable anisotropic behavior, drawing inspiration from bio-architectures.

Throughout billions of years of evolution, nature has introduced several optimized architectures with extreme performances in many creatures, sustaining complex circumstances and loading situations[47,48]. Taking inspiration from nature, this study combines the concepts of nesting, the golden spiral, fractals, and three four-fold axis of rotational symmetry to realise novel lattices and delve into the expansive design space. Nesting architectures comprise structure within structure[49] and are found in cross sections of many bio-architectures such as cortical bone osteons[50] and trees[50] providing the "nesting order" concept[43]. The majority of the plants showcase specific orientations, such as the golden spiral[51–53] and repeating patterns of self-similar architectures known as fractals[54,55]. Drawing inspiration from golden spirals and fractals, the "nesting orientations" and "self-similar repetition of X-cross struts" are implemented. Additionally, the concept of three four-fold axes of rotational symmetry is also incorporated to maintain the cubic symmetry of lattice structures[56].

This study proposes bio-inspired mono-, bi- and tri-nest lattices considering aspects such as NOs, NORs, X-cross struts repetitions and cubic symmetry. The diverse design attributes of these lattice structures are systematically varied to explore the vast design space. Finite element analysis in conjunction with numerical homogenization technique is utilized to evaluate the stiffness tensor for each of the unit cell architectures. Subsequently, anisotropic measure - the Zener ratio Z, the normalized Young's modulus $\overline{E}$, the relative density $\overline{\rho}$ and the surface area density $\overline{S}$ are evaluated. The anisotropic behavior of lattices was analysed to identify optimal ILSs. Analyses indicate that for mono-nest lattices, as NO increases, the anisotropic behavior transitions from shear dominant (SD) behavior to tensile/compression dominant (TCD) behavior for a given NOR and at the same time, for a given NO as the NOR angle increases, the behavior transitions from TCD to isotropic behavior. These transitions in anisotropic behavior are considered as foundations for generating a greater number of ILSs which can be beneficial for many structural applications. Additionally, most of the multi-nest lattices are also found to exhibit perfectly isotropic and neo-isotropic behavior.

## 2. Methodology

### 2.1. Bio-inspired nested lattice structures (NLSs)

In the natural structure of both animal and human cortical bones, one finds a combination of primary bone and osteons. Fig. 1a illustrates a cross-section of the cortical bone in the femur. The cortical bone's osteons[50], depicted in Fig. 1a, encompass central haversian canals, encircled by multiple concentric lamellae arranged in a cylindrical fashion. These concentric arrangements adhere to a nesting phenomenon, exemplifying a structure within a structure[50]. This nesting phenomenon serves as a foundational concept in this study, guiding the development of nested lattice structures (NLSs) derived from the intricately inscribed nested cubical structures within nested osteon rings, as illustrated in Fig. 1a. Each of these nested osteon rings is visually distinguished by colors green, pink and sky blue, and their corresponding inscribed nested cubical structures are similarly indicated with these hues. Drawing inspiration from these biological structures, we define three distinct NOs.

Nature also possesses an innate ability to craft intricate and aesthetically pleasing structures that also deliver remarkable functionality. Fractal geometry stands out as a prime example. Fractals are characterized by the recurring presence of self-similar geometric shapes or patterns across varying length scales[54]. They are prevalent in the natural world, as depicted in Fig. 1b, showcasing examples such as pineapples, succulents (echeveria), and aloe polyphylla (left to right). The majority of plant species demonstrate spiral growth patterns, a phenomenon elegantly elucidated by the renowned Fibonacci sequence, also known as the golden spiral[51,52]. This research draws inspiration from these two fundamental principles, namely fractals and the golden spiral, to generate nested cubical lattice structures at varying orientations by rotating them within the nested osteon rings while preserving their inherent architecture across all nesting orders. Based on these biological inspirations, three distinct NORs in relation to NOs are considered. Additionally, the concept of cubic symmetry i.e., the Three-Four-Fold Axis of Rotational Symmetry (T4FAS), is considered as a framework for designing symmetric NLSs. T4FAS entails the ability of lattice structures to return to their original configuration after a 90 degrees rotation about a designated axis as depicted in pink in Fig. 1c. All the above steps are carried out sequentially, in conjunction with T4FAS, to produce the final symmetric NLSs as shown in Fig. 1d.

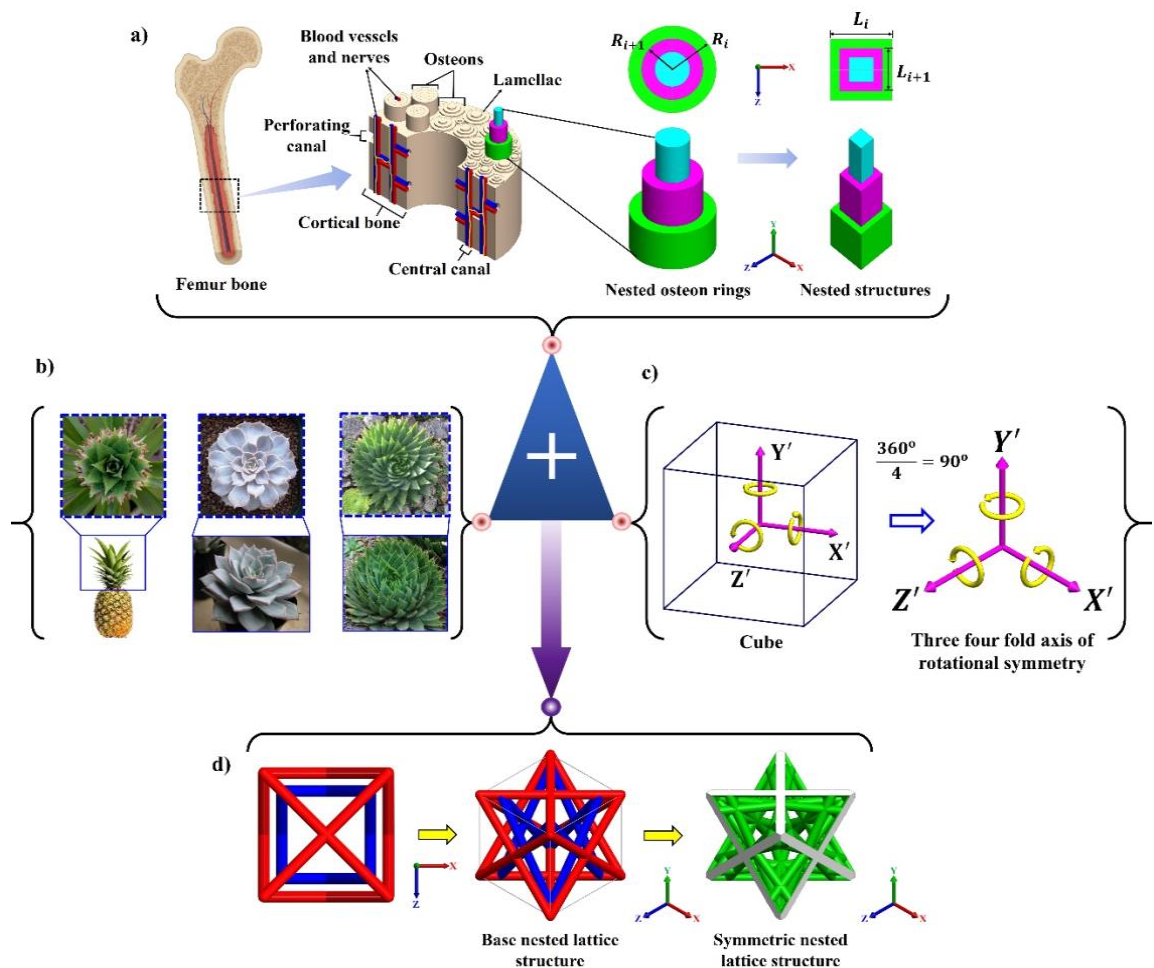

Fig. 1. Schematic representation of proposed methodology for developing lattice materials a) bioinspired nesting orders (NOs) mimicking the architecture of cortical bone osteons, b) nesting orientations (NORs) inspired by the fractals and golden spiral, c) three four-fold axis of rotational symmetry (T4FAS) to ensure cubic symmetry and d) 3D nested-lattice structures (NLSs).

### 2.1.1. Theoretical form of base XNLSs

Fig. 2 illustrates the generalized theoretical form of the base NLS, which represents the structures before the application of T4FAS. This base structure is formed by combining NOs and their respective NORs, in addition to the self-similar structures of X-cross struts. The creation of this base NLS involves several steps. As depicted in Fig. 2a, the inscribed nested squares, indicating NOs i.e., $N_i$ (red square) and $N_{i+1}$ (blue square), are generated by drawing inspiration from nested osteon rings. These rings have radii $R_i$ and $R_{i+1}$ respectively. Here, i = 0, 1, 2...n, and are separated by a distance of $\alpha_i$. The side length of squares i.e., NOs are denoted by $L_i$ and $L_{i+1}$, and their respective NORs by $\theta_i$ and $\theta_{i+1}$ respectively. Except $N_i$, which remains fixed with $\theta_i$ at $0^o$ to preserve the cubic shape, the remaining NOs are oriented at varying angles in an anti-clockwise direction relative to the diagonal kk, which coincides with the $Z''$-axis.

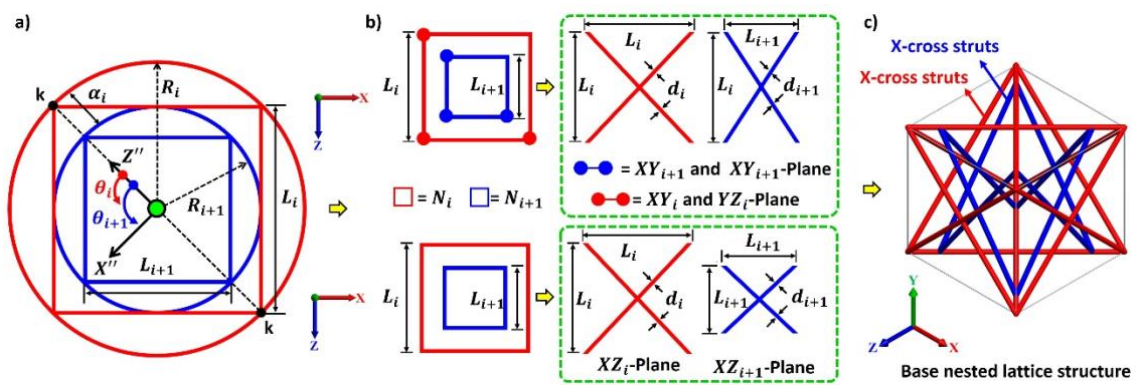

Fig. 2. a) NOs and NORs within nested circles, b) X-cross struts on XY, YZ, and XZ planes of respective NOs and NORs, and c) 3D view of X-cross struts within the nested base lattice structure (only two NOs are shown for brevity).

Drawing inspiration from the concept of fractals, this study incorporates X-cross struts, positioning them on the XY and YZ planes of their respective NOs as depicted in Fig. 2b. Likewise, these X-cross struts are also situated on the XZ plane. Subsequently, each cubical NO incorporates X-cross struts within the NLS, as illustrated in Fig. 2c. The diameters of these X-cross struts for the respective NOs are denoted as $d_i$ and $d_{i+1}$. Following the assignment of these diameters to the corresponding X-cross struts, the base X-cross NLS (XNLS) is formed, as depicted in Fig. 2c. Utilizing this base XNLS as a foundation, the symmetric XNLS is produced following the application of T4FAS, highlighted in green (see, Fig 1d). Note that only two NOs are shown in Fig. 2c for brevity.

A mathematical relationship governing the geometrical parameters of the base XNLS is established, as given by the following Eq. (1). Here, $L_{i+1}$ is expressed as a function of the side length of the preceding NO, i.e., $L_i$, and the separation distance between the two nested osteon rings, $\alpha_i$. It is important to note that $\alpha_i$ is treated as a constant in this investigation.

$$L_{i+1} = \sqrt{2}\left[\left(\frac{L_i}{\sqrt{2}}\right) - \alpha_i\right]; \; i = 0, \; 1, \; 2, \ldots n \tag{1}$$

By considering geometrical, physical and manufacturing constraints, this study is limited to three NOs: $N_0$, $N_1$ and $N_2$, and their respective NORs are $\theta_0$, $\theta_1$ and $\theta_2$ where $\theta_1$ and $\theta_2$ vary from $0^o$ - $45^o$ with the step of $15^o$ and $\theta_0$ is fixed at $0^o$ due to the geometrical constraints i.e., to maintain the cubic shape. Here $L_0$, is considered as 8.75mm, the $\alpha_0$ and $\alpha_1$ are considered as 1.81mm due to physical and manufacturing constraints, and the strut diameters $d_0$, $d_1$, and $d_2$ are considered as 0.8mm due to constraints of manufacturing. For comparison of different XNLSs and to conduct the parametric evaluations, these diameters are adjusted based on the relative density ($\bar{\rho}$). All the XNLSs are designed using SolidWorks.

Finally, in the pursuit of isotropic XNLSs, we first explore mono-nest architectures using specified NOs and their corresponding NORs. Subsequently, we delve into designing multi-XNLSs by combining NOs such as $(N_0, N_1)$ and $(N_0, N_1, N_2)$ with their respective NORs, which include $(\theta_0, \theta_1)$ and $(\theta_0, \theta_1, \theta_2)$ respectively. The subsequent sections thoroughly discuss the design processes and architectures of these XNLSs.

### 2.2. Geometric Modeling of XNLSs

The construction process of XNLSs encompasses various steps, outlined as follows:

*Step I*: Initially, NOs of $N_0$ and $N_1$ are considered, along with their respective NORs, $\theta_0$ and $\theta_1$, both set at $0^o$. Strut diameters of 0.8mm are assigned. Next, a four-fold operation is carried out, which entails orienting the base XNLS about the Y-axis at $90^o$ intervals up to $360^o$, resulting in four base XNLS placed at $90^o$ intervals around the Y-axis and the union formed the structures. The structures formed from this operation are shown in yellow.

*Step II*: The NLS folded along the Y-axis exhibits symmetry solely about the Y-axis, lacking symmetry about the other axes. The T4FAS operation is subsequently applied to this Y-axis folded NLS, involving a four-fold operation along the X-axis. The resultant structures are then consolidated, and the newly formed struts are highlighted in blue, illustrating their integration with the Y-axis folded NLS. Similarly, the NLS folded along the X-axis displays symmetry exclusively along the X-axis, with no symmetry observed along the other axes.

*Step III*: To fully complete the T4FAS operation and establish cubic symmetry within the lattice structure, the X-axis folded NLS undergoes a four-fold operation along the Z-axis. The resulting structures are combined, and the newly formed struts are indicated in green, indicating their degree of merging with the preceding NLS. This process ensures symmetry across all three axes, meeting the criteria for cubic symmetry. Finally, any surplus struts beyond the unit cell bounding box are eliminated through Boolean operations, ultimately producing the unit cell architecture of symmetric XNLS. This construction process is delineated in Supplementary Video, SV. This methodology is consistently applied to other XNLSs in this study.

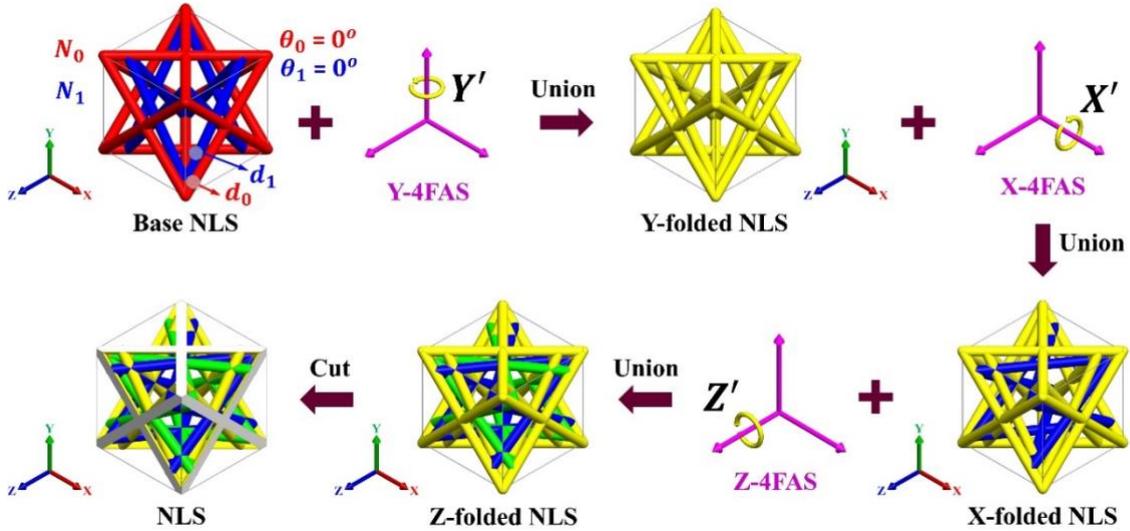

Fig. 3. An overview of the steps involved in the design of a unit cell geometry of XNLSs.

## 2.3. Bio-inspired 3D XNLSs

Here, we categorize XNLSs into two distinct groups. The initial category of XNLSs comprises mono-nest, wherein the X-cross struts are positioned within a single NO and various NORs within the NO contribute to the formation of the Mono-XNLSs. The second category involves the amalgamation of different NOs and their corresponding NORs, with the X-cross struts placed within them to generate the Multi-XNLSs. The subsequent sections provide a detailed delineation of these two classes.

### 2.3.1. Mono-XNLSs

The Mono-XNLSs of respective NOs and NORs are shown in Fig. 4. Each row of Fig. 4a represents the different NOs, and each column represents the NORs. Henceforth, we denote them as "XNOO: NOs: NORs". In this notation, XNOO indicates the X-cross nesting order and orientation. NOs are written as 0, 1 and 2, and NORs are written as 0, 15, 30 and 45. XNOO:0:$\theta_0$, XNOO:1:$\theta_1$ and XNOO:2:$\theta_2$ are colored red, blue, and green (see, Fig. 4a).

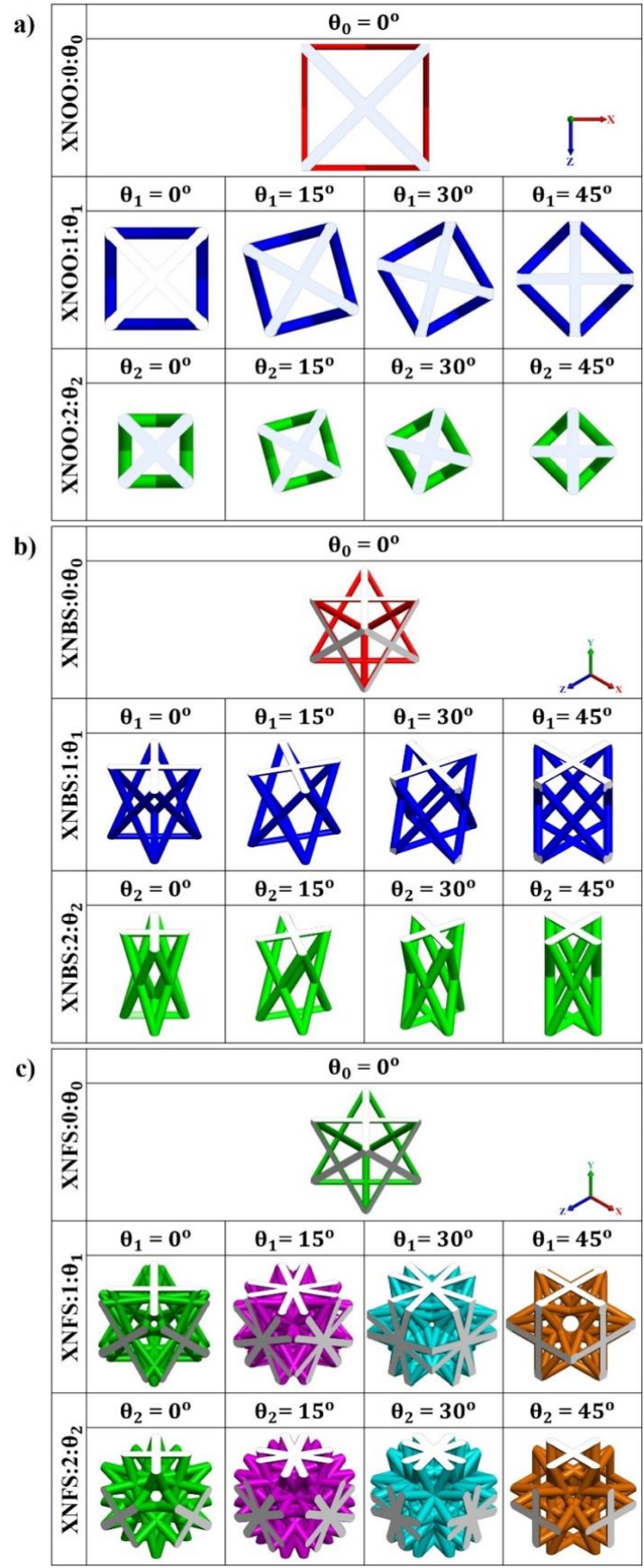

Fig. 4. Mono-XNLSs considering different NOs and NORs: a) XNOOs b) XNBSs and c) XNFSs

Their respective X-cross nested base structures (XNBSs) are shown in Fig. 4b. We also denote them as XNBS:0:$\theta_0$, XNBS:1:$\theta_1$ and XNBS:2:$\theta_2$ and are shown in red, blue and green respectively. By considering these XNBSs, the X-cross nested full structures (XNFSs) are generated carrying out the T4FAS operation. These XNFSs are also called as Mono-XNLSs and are shown in Fig. 4c. The full structures are denoted as XNFS:0:$\theta_0$, XNFS:1:$\theta_1$ and XNFS:2:$\theta_2$ where $\theta_0$, $\theta_1$ and $\theta_2$ denote their respective orientation angles ($\theta_0 = 0°$; $\theta_1$ and $\theta_2$ = 0°, 15°, 30° and 45°). In this study, a total of 9 Mono-XNLSs are generated and they are shown in different colors to distinguish them from each other.

### 2.3.2. Multi-XNLSs

2.3.2.1. Bi-XNLSs considering $N_0$ and $N_1$ nesting orders

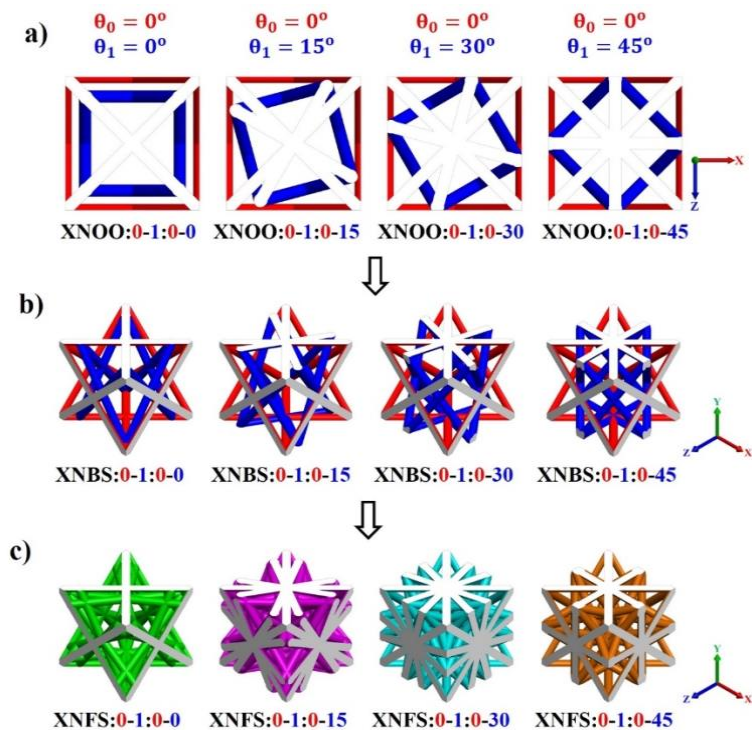

Fig. 5. Bi-XNLSs considering $N_0$ and $N_1$: a) XNOOs b) XNBSs and c) XNFSs

The NOs, $N_0$ and $N_1$ along with their respective NORs, $\theta_0$ and $\theta_1$, are considered to generate bi-nest lattices, referred to as Bi-XNLSs, as illustrated in Fig. 5. The corresponding XNOOs are presented in Fig. 5a. It is important to note that the first nesting orientation is fixed ($\theta_0 = 0°$), depicted in red, while $\theta_1$ is varied from 0° to 45° in increments of 15°, represented in blue. The XNOOs serve as the basis for generating XNBSs, and upon applying the T4FAS operation (as shown in Fig. 5b, c), XNFSs are obtained. The names of these structures are indicated in Fig. 5. Taking the example of the XNLS labelled as XNFS:0-1:0-0, where "0-1" signifies the combination of NOs of $N_0$ and $N_1$ "0-0" represents the combination of NORs of $\theta_0$ and $\theta_1$. Similar indications apply to XNOOs and XNBSs. A total of 4 Bi-XNLSs are designed by combining $N_0$ and $N_1$, and respective NORs.

2.3.2.2. Tri-XNLSs considering $N_0$, $N_1$ and $N_2$ nesting orders

Here, we consider three different NOs, $N_0$, $N_1$ and $N_2$, along with their corresponding NORs ($\theta_0$, $\theta_1$ and $\theta_2$) to realise tri-nest unit cell architectures. These lattices are also referred to as Tri-XNLSs. The NOR of $\theta_0$ is set at $0^o$ to maintain the cubic shape, while the remaining NORs of $\theta_1$ and $\theta_2$ are varied from $0^o$ to $45^o$ in $15^o$ increments. The resulting XNOOs of the combined Tri-XNLSs are illustrated in Fig. 6a, with the NOs colored in red, blue, and green for clarity. Subsequently, XNBSs are generated based on these XNOOs, as depicted in Fig. 6b, followed by XNFS, which are formed after the application of T4FAS operation. These XNLSs are now denoted as "XNFS: NORs", as shown in Fig. 6c, with a total of 16 XNLSs designed by combining the three NOs and their respective NORs. The Tri-XNFSs are differentiated by distinct colors.

A total of 9 Mono-XNLSs and 20 Multi-XNLSs (4 Bi-XNLSs and 16 Tri-XNLSs) are created to achieve isotropic XNLSs with tuneable anisotropic behavior. Here, we focus on the geometric characteristics of XNLSs, including relative density $\bar{\rho}$, defined as the ratio of XNLS volume to the bulk unit cell volume of XNLS, and surface area-to-volume ratio $\bar{S}$, representing the ratio of XNLS's surface area to volume. Additionally, the elastic mechanical properties of the XNLSs are evaluated, and these aspects are further detailed in the next section.

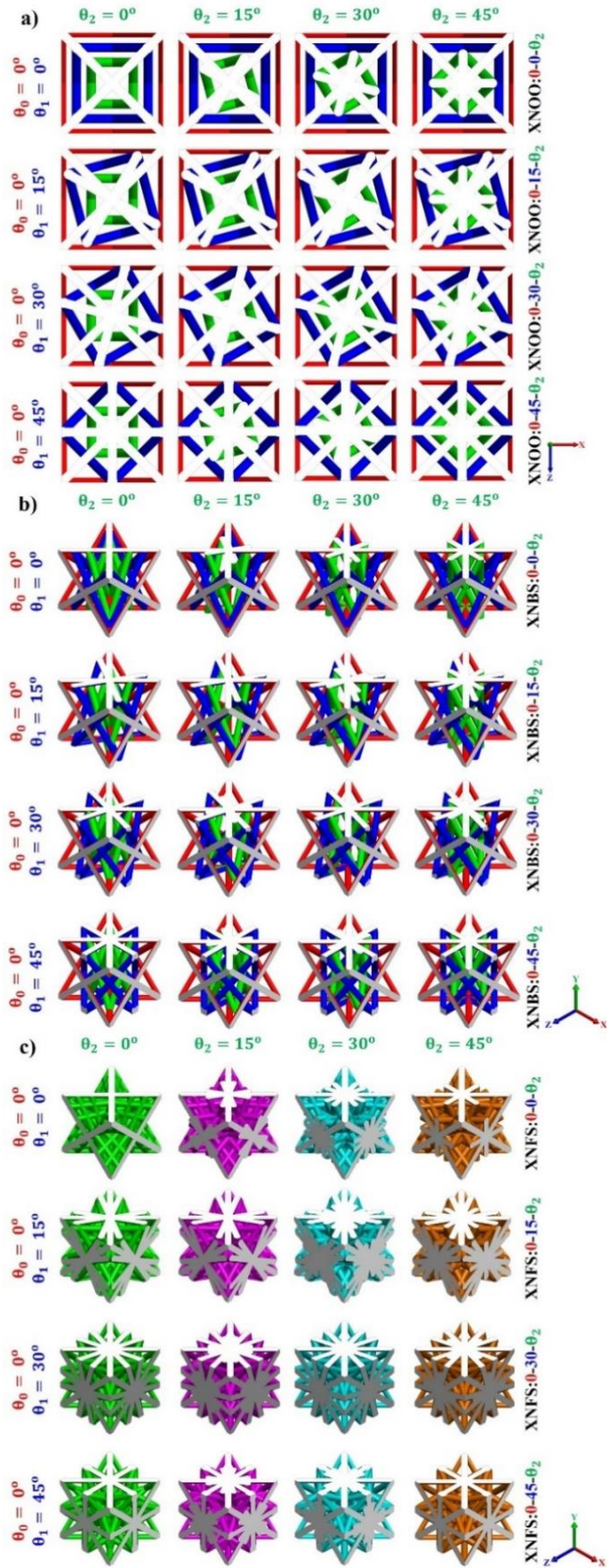

Fig. 6.  Tri-XNLSs considering $N_0$, $N_1$ and $N_2$: a) XNOOs b) XNBSs and c) XNFSs

## 2.4. Numerical homogenization method

The numerical homogenization technique proposed by the Steven[57] is utilised to evaluate the elastic anisotropic continuum mechanical properties of non-continuum periodic lattice structures. As per the generalized Hooke's law, this can be expressed as[43]:

$$\boldsymbol{\sigma} = \boldsymbol{C} \cdot \boldsymbol{\varepsilon} \qquad (2)$$

The second order stress tensor $\boldsymbol{\sigma}$ is connected to second order strain tensor $\boldsymbol{\varepsilon}$ through a fourth order stiffness tensor i.e., $\boldsymbol{C}$ which contains 81 independent elastic constants and elastic constants are reduced to 21 considering inherent symmetries of $\boldsymbol{C}$, $\boldsymbol{\sigma}$ and $\boldsymbol{\varepsilon}$[45]. The fourth order tensor $\boldsymbol{C}$ defines the elastic anisotropic characteristics of the periodic lattice structures made of a given material. For isotropic materials, only two independent elastic constants are required and they are $C_{11}$ and $C_{12}$. This study considered the symmetric cubic system where $C_{11} = C_{22} = C_{33}$, $C_{12} = C_{23} = C_{13}$ and $C_{44} = C_{55} = C_{66}$, and the remaining constants are zero, as expressed in the Eq. (3). The elements of the tensor govern the anisotropic behavior of the XNLSs and unit cell architectures can be tuned to design the isotropic XNLSs.

$$C_{ij} = \begin{bmatrix} C_{11} & C_{12} & C_{12} & 0 & 0 & 0 \\ C_{12} & C_{11} & C_{12} & 0 & 0 & 0 \\ C_{12} & C_{12} & C_{11} & 0 & 0 & 0 \\ 0 & 0 & 0 & C_{44} & 0 & 0 \\ 0 & 0 & 0 & 0 & C_{44} & 0 \\ 0 & 0 & 0 & 0 & 0 & C_{44} \end{bmatrix} \qquad (3)$$

The Young's modulus E, and anisotropic index, also known as Zener ratio Z, are related to the elements of the tensor $C_{ij}$ as expressed in Eq. (4) and (5) respectively. Note that they are evaluated from three independent elastic constants $C_{11}$, $C_{12}$ and $C_{44}$ to characterise the mechanical performance of XNLSs[39].

$$E = \frac{(C_{11}^3 + 2C_{12}^3 - 3C_{11}C_{12}^2)}{(C_{11}^2 - C_{12}^2)} \qquad (4)$$

$$Z = \frac{C_{44}}{\left[\dfrac{C_{11} - C_{12}}{2}\right]} \qquad (5)$$

The anisotropic behavior of XNLSs is evaluated based on 3D Young's modulus surface (3DYMS) which is independent of the base material[44]. If the 3DYMS has a spherical shape then XNLS is considered isotropic, where $Z = 1$, otherwise anisotropic. If the shape of the 3DYMS for a given unit cell architecture has protrusion along the axes then the lattice is considered to exhibit tensile/compression dominant (TCD) behavior, where $Z < 1$ and if the shape has concavity on the faces of the cube, then it is considered as shear dominant behavior where $Z > 1$[39,44].

The normalized elastic modulus of the XNLSs, $\bar{E}$ is calculated by dividing the modulus of the cellular structure, $E$ with that of the base material, $E_s$ as given in the Eq. (6).

$$\bar{E} = \frac{E}{E_s} \qquad (6)$$

To obtain the $C_{ij}$ tensor, nTopology tool is utilised to perform the numerical homogenization imposing periodic boundary conditions. In this tool, the "Homogenization of unit cell" algorithm is used along with defined "bounding box", which is considered as design volume, for a given XNLS. The properties of parent material, 316L-Stainless Steel (316L SS) are listed in the Table S1. Quadratic tetrahedral mesh elements with 0.3mm size are used for all XNLSs to ensure accuracy of results. The details of the validation of this finite element analysis are given in Supporting Information (see S1).

## 3. Results and discussion

A range of bio-inspired Mono-, Bi-, and Tri-nest lattices were constructed and the finite element analysis (FEA) in conjunction with numerical homogenization technique was utilized to assess the performance attributes of XNLSs. The assessment included the evaluation of geometrical properties ($\bar{\rho}$ and $\bar{S}$) and elastic properties ($Z$, 3DYMSs and $\bar{E}$). These parameters were employed to characterize various XNLSs based on NOs and respective NORs. The subsequent sections provide a comprehensive discussion on the influence of NOs and NORs on the performance attributes of these XNLSs.

### 3.1. Mono-XNLSs

This section discusses the effect of different individual NOs and respective NORs on the performance characteristics of Mono-XNLSs which are shown in Fig. 7. Each cell (box) of Fig. 7a contains 3DYMSs, $Z$ and $\bar{E}$. Each row represents the NOs ($N_0$, $N_1$ and $N_2$) and each column indicates the NOR from $0^o$ to $45^o$.

According to Fig. 4, as the NO is varied from $N_0$ to $N_2$ for a fixed NOR of $0^o$, the number of struts and nodes, strut connectivity and heterogeneity of the porous architecture increase. The XNFS:0:0 has less number of half cross-section struts at the boundary of the unit cell. Similarly, as $\theta_1$ and $\theta_2$ are increased from $0^o$ to $45^o$, resulting XNLSs also evince higher number of struts and nodes with more strut connectivity, especially in the range of $15^o$ to $30^o$. Overall, both NOs and NORs have more influence on the architectures of Mono-XNLSs.

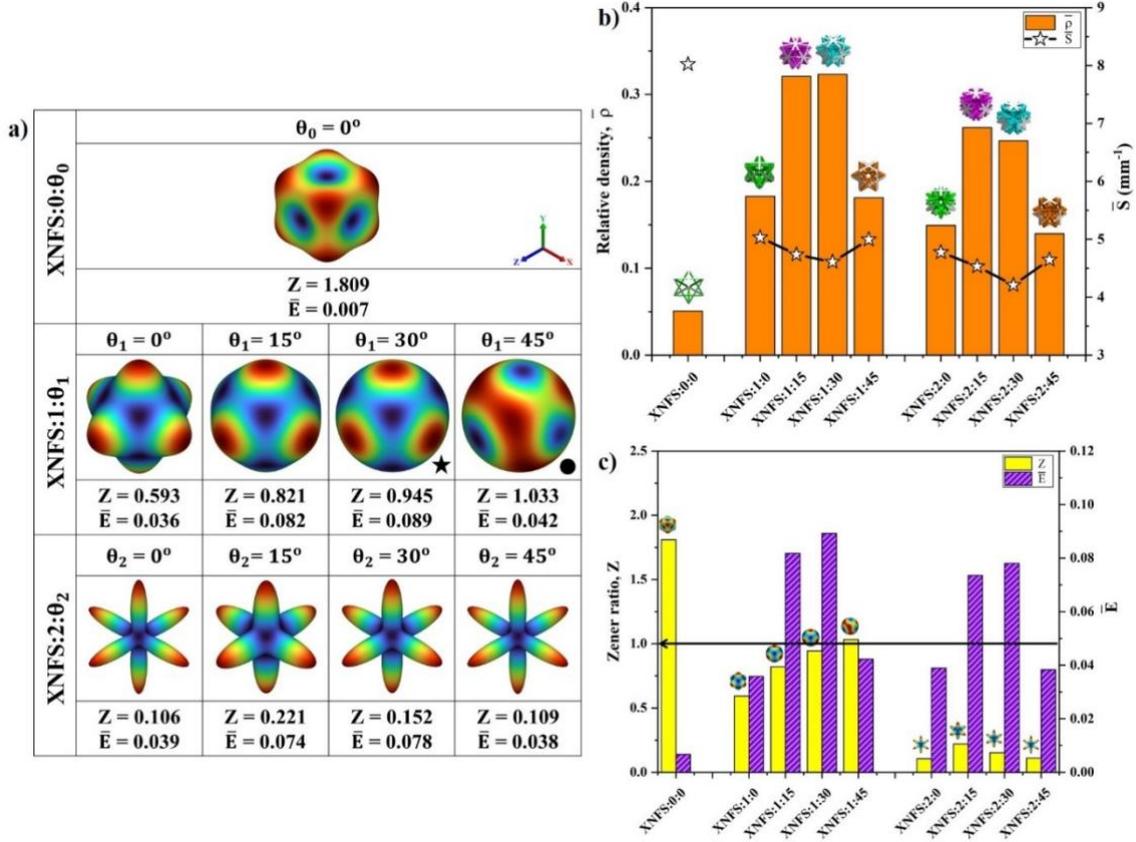

Fig. 7. Performance of Mono-XNLSs: a) 3DYMSs, b) $\bar{\rho}$ and $\bar{S}$, and c) Z and $\bar{E}$

From Fig. 7b, it is observed that both NOs and NORs have a significant influence on $\bar{\rho}$ and $\bar{S}$. For NOR of $0^o$, as the NO increased from $N_0$ to $N_2$, the $\bar{\rho}$ increases at a higher rate and the $\bar{S}$ drastically reduces from $N_0$ to $N_1$ due to the presence of half cross-section struts in XNFS:0:0 architecture. Then $\bar{\rho}$ gradually decreases and $\bar{S}$ further reduces at a lower rate from $N_1$ to $N_2$. Similar observations are found for other $\theta_1$ to $\theta_2$. Similarly, at a given $N_1$, as $\theta_1$ is increased from $0^o$ to $45^o$, the $\bar{\rho}$ and $\bar{S}$ greatly vary, where, from $0^o$ to $15^o$, $\bar{\rho}$ increases at a higher rate due to the presence of more number of struts in XNFS:1:15 compared with XNFS:1:0. Then from $15^o$ to $30^o$ very small changes in $\bar{\rho}$ observed due to XNFS:1:15 and XNFS:1:30 having almost the same number of struts, and finally for $30^o$ to $45^o$, $\bar{\rho}$ gradually reduces with a higher rate due to decrease in number of struts. Similar observations are found for XNLSs of $N_2$ at different $\theta_2$. Based on these results, $N_1$-based Mono-XNLSs have higher $\bar{\rho}$ followed by $N_2$- and $N_0$-based structures. Within given $N_1$ to $N_2$, both $\theta_1$ to $\theta_2$ of $15^o$ to $30^o$ exhibit highest $\bar{\rho}$ compared with other angles of $0^o$ to $45^o$. A wide range of $\bar{\rho}$ between 0.051 and 0.323 is observed. The XNFS:0:0 has the highest $\bar{S}$ due to the presence of half struts at the boundaries of unit cell. As both $\theta_1$ and $\theta_2$ are increased from $0^o$ to $45^o$, the changes observed in $\overline{S}$ are insignificant, but both exhibit almost similar trend.

The 3DYMSs of these Mono-XNLSs are shown in Fig. 7a. We observed the transition of SD to TCD anisotropic behavior as the NO changed from $N_0$ to $N_2$ for NOR of $0^o$ where their Z values decreased from 1.809 to 0.106. This decrease is high for $N_0$ to $N_1$, i.e., from 1.809 to 0.593, compared with $N_1$ to $N_2$ for which Z varies from 0.593 to 0.106. Similarly, for other NORs of $\theta_1$ and $\theta_2$, from $15^o$ to $45^o$, as the NO is increased from $N_1$ to $N_2$, the Z value decreases. As anisotropic behavior is changing from SD to TCD while the NO is increased from $N_0$ to $N_2$ at a given NOR of $0^o$ with equal spacing of $\alpha =1.81$mm, the anisotropic measure Z is expressed in Eq. (7) as a function of the ratio of spacing between nesting orders $\alpha$ and unit cell size $L_0$ where coefficients $a$, $b$ and $c$ are constants and they depend on the unit cell architecture. Note the isotropic XNLSs can be achieved between the NOs of $N_0$ and $N_1$ due to transition from SD to TCD.

$$Z(\bar{\alpha}) = a\bar{\alpha}^2 + b\bar{\alpha} + c ; \qquad R^2 = 1 \qquad (7)$$

Here $\bar{\alpha} = \frac{\alpha}{L_0}$. Additionally, it is also found that as the $\theta_1$ increased from $0^o$ to $45^o$, the anisotropic behavior of XNLSs changed from TCD to isotropic behavior where the Z value increases from 0.593 to 1.033. This transition has a polynomial relation between Z and $\theta_1$ and is given by

$$Z(\theta_1) = p\theta_1^3 + q\theta_1^2 + r\theta_1 + s; \qquad R^2 = 1 \qquad (8)$$

Here, $p$, $q$, $r$, and $s$ are constants which depend on NOR of the unit cell architecture. The XNFS:1:30 and XNFS:1:45 are indicated by a dark star and circle which are considered as neo-isotropic (Z = 0.900 to 0.950 and 1.050 to 1.100) and perfectly isotropic (0.950 < Z < 1.050) XNLSs[38]. The 3DYMSs of $N_1$-based Mono-XNLSs transform from protruded shapes along axes into a spherical shape. But $N_2$-based Mono-XNLSs exhibit no such transition and all of them exhibit TCD behavior and their 3DYMSs resemble needle-like shapes along the axes.

In Fig. 7c, the horizontal arrow represents the isotropic behavior (i.e., Z = 1). The Z values of XNFS:0:0 lie above the arrow (i.e., Z > 1) indicating SD behavior. The Z values of XNFS:1:30 and XNFS:1:45 situated closer to the line are considered neo-isotropic and perfectly isotropic XNLSs, while the rest of the XNLSs are positioned below the arrow, indicating TCD behavior. From these observations, it can be summarized that as more NOs are included in the cubic unit cell of XNLSs with respect to NORs, the anisotropic behavior becomes highly TCD transitioning from SD. This is due to the size of cubical NO becoming smaller and with increasing NO, the XNLSs attain rod-like geometry along axes and exhibit high TCD behavior. Furthermore, within this transition, we observe isotropic XNLSs at $N_1$ and $\theta_1$ of 45°. Based on Z values and 3DYMSs of Mono-XNLSs, it is found that the NO has more influence on the anisotropic behavior than NOR.

From Fig. 7c, it can be seen that both NOs and NORs have a high influence on $\overline{E}$. As NO increases, $\overline{E}$ increases initially at a higher rate and then gradually increases for a given NOR of 0°. Also, as $\theta_1$ increases from 0° to 45°, the XNFS:1:15 and XNFS:1:30 exhibit higher $\overline{E}$ compared with other architectures. Similarly, $N_2$-based Mono-XNLSs also follow the same trend. Overall, $N_1$-based Mono-XNLSs exhibited the highest $\overline{E}$ and followed by $N_2$- and $N_0$- based Mono-XNLSs. Based on these observations, both NOs and NORs have an impact on $\overline{E}$. However, it is noticed that NORs have more influence on $\overline{E}$ than NOs. All the performance attributes of these Mono-XNLSs are highly influenced by both NOs and NORs. Very interestingly, $\theta_1$ and $\theta_2$ of 0° and 45°-based Mono-XNLSs can be much more beneficial for the load-bearing bone implant applications which require low $\overline{\rho}$, high $\overline{S}$, matching $\overline{E}$ and heterogenous pores[39,45] which are achieved by these bioinspired Mono-XNLSs. So far, the strut diameter is kept at 0.8mm. The impact of varying strut diameters (ranging from 0.6 to 1.0mm with 0.1mm increments) on the performance of Mono-XNLSs is detailed in Supporting Information S2, revealing significant influence on $\overline{\rho}$, $\overline{S}$ and $\overline{E}$, across specified NOs and NORs, while demonstrating minimal effect on anisotropic behavior[39].

## 3.2. Bi- and Tri-XNLSs

### 3.2.1. Bi-XNLSs

Upon observing the anisotropic behavior of Mono-XNLSs, we expanded the bioinspired design process by incorporating two NOs to produce XNLSs and tune their anisotropic properties. The 3DYMSs of Mono-XNLSs, originating from NOs $N_0$ and $N_1$, oppose each other in terms of anisotropic behavior. Specifically, while XNLSs of $N_0$ exhibit SD behavior, those of $N_1$ primarily demonstrate TCD behavior, except for XNFS:1:45, which exhibits perfectly isotropic behavior. By synergistically combining these two families of lattices with contrasting anisotropic behaviors through a method known as hybridization to counter 3DYMSs[38,39,44], a variety of XNLSs are realized, which exhibit isotropy. The designs of combined XNBSs and XNFSs from both $N_0$ and $N_1$, along with their respective NORs, are depicted in Fig. 5.

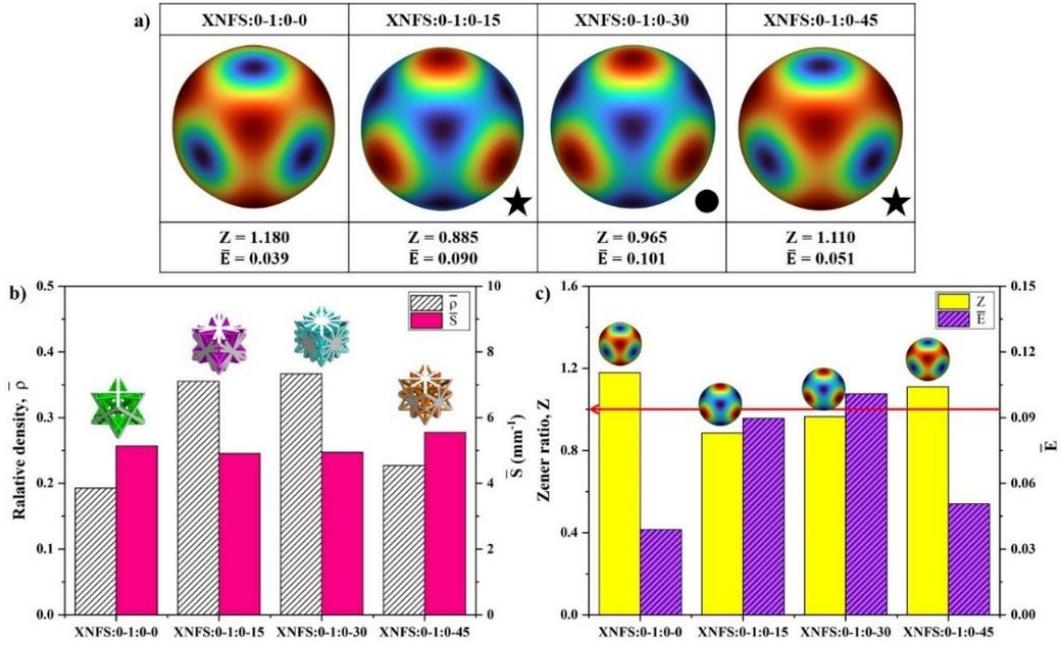

Fig. 8. The performance of Bi-XNLSs: a) 3DYMSs, b) $\bar{\rho}$ and $\bar{S}$, and c) Z and $\bar{E}$

Based on Fig. 5c, the unit cell architectures of Bi-XNLSs differ significantly from each other and from Mono-XNLSs. They exhibit higher strut connectivity, more struts and nodes, and heterogeneous pores. From Fig. 8b, $\bar{\rho}$ values of both XNFS:0-1:0-15 and XNFS:0-1:0-30 are higher due to more struts, while XNFS:0-1:0-0 and XNFS:0-1:0-45 have low and similar $\bar{\rho}$ values. The $\bar{\rho}$ values of these Bi-XNLSs range from 0.193 to 0.367. The $\bar{S}$ of these Bi-XNLSs ranges from 4.910 to 5.549 mm$^{-1}$, showing little variation. In Fig. 8c, $\bar{E}$ initially increases from XNFS:0-1:0-0 to XNFS:0-1:0-15, then slightly increases from XNFS:0-1:0-15 to XNFS:0-1:0-30, and finally reduces from XNFS:0-1:0-30 to XNFS:0-1:0-45. Both XNFS:0-1:0-15 and XNFS:0-1:0-30 have higher $\bar{E}$ values compared to other architectures, ranging from 0.039 to 0.101.

Interestingly, the Z values of Bi-XNLSs range from 0.885 to 1.180, indicating closer to unity and almost spherical 3DYMSs as shown in Fig. 8a. All Bi-XNLSs align close to the red arrow in Fig. 8c, while XNFS:0-1:0-30 exhibits a Z value of unity, demonstrating isotropic behavior. The Z values of XNFS:0-1:0-15 and XNFS:0-1:0-45 are also close to unity, indicating neo-isotropic behavior. In Fig. 8a, the 3DYMS of XNFS:0-1:0-0 is almost spherical, exhibiting SD behavior, whereas its Mono-XNLSs display both SD and TCD behaviors (see Fig. 7a). This suggests that the first NO and its NORs have more influence on Z than those of the second NO. However, in the case of XNFS:0-1:0-15 and XNFS:0-1:0-30, this influence is less pronounced. XNFS:0-1:0-45 shows neo-isotropic behavior due to being a combination of Mono-XNLSs with SD and isotropic behaviors.

### 3.2.1.1. Bi-XNLSs for prescribed relative densities

To compare the performances of Bi-XNLSs, we designed them for a fixed $\bar{\rho}$ of 0.10 by altering the strut diameters while keeping the same diameter for both NOs and maintaining a constant unit cell size.

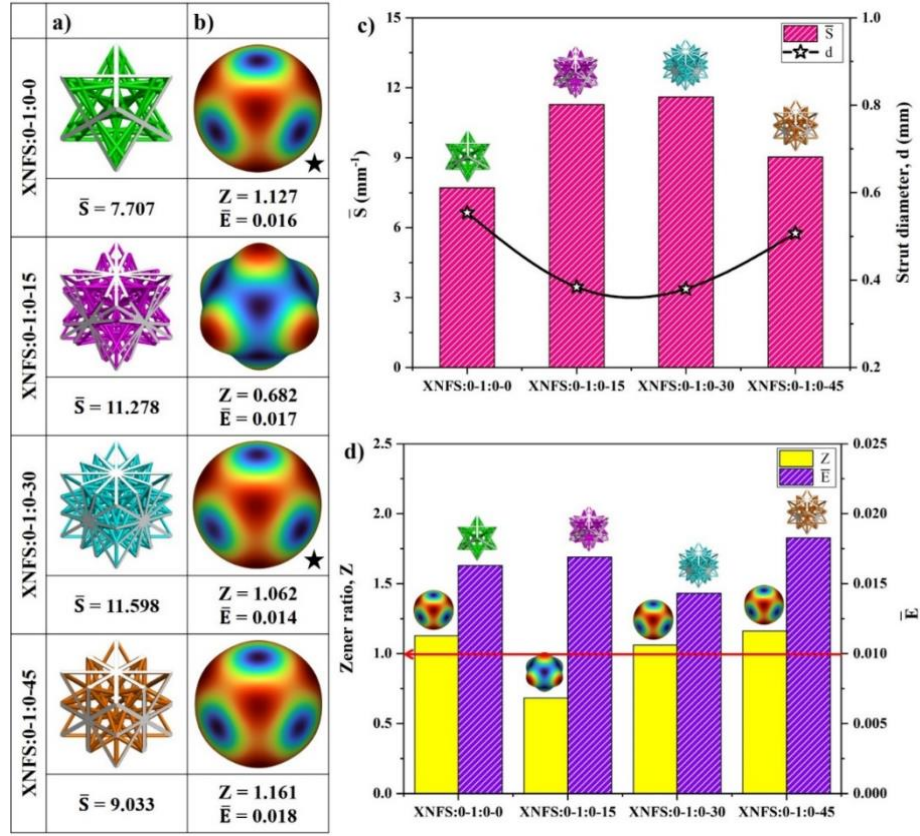

Fig. 9. Performance attributes of Bi-XNLSs at $\bar{\rho} = 0.10$: a) unit cell architectures, b) 3DYMSs, c) $\bar{S}$ and variation in strut diameter, and d) Z and $\bar{E}$

The diameter of the struts needs adjustment within the range of 0.375 to 0.385mm for the Bi-XNLSs shown in Fig. 9c to maintain a $\bar{\rho}$ of 0.10. As anticipated, the trend of the $\bar{S}$ is inversely related to the strut diameter. It increases from XNFS:0-1:0-0 (7.707mm$^{-1}$) to XNFS:0-1:0-15 (11.278mm$^{-1}$), then slightly rises from XNFS:0-1:0-15 to XNFS:0-1:0-30 (11.598mm$^{-1}$), and subsequently decreases from XNFS:0-1:0-30 to XNFS:0-1:0-45 (9.033mm$^{-1}$). In the context of fixed unit cell size strut-based lattice structures, a reduction in strut diameter increases $\bar{S}$, whereas in surface-based lattices, the opposite occurs[58]. Notably, there's a wider range of $\bar{S}$ observed here, spanning from 7.707 to 11.598mm$^{-1}$, where XNFS:0-1:0-30 and XNFS:0-1:0-15 exhibit the highest $\bar{S}$ compared to others, attributed to nesting, which could benefit load-bearing bone cellular activities such as cell adhesion, growth, proliferation, and differentiation[40].

From Fig. 9b and 9d, the Z remains close to unity. Due to lowest $\bar{\rho}$, there is minimal variation in Z. While XNFS:0-1:0-15 and XNFS:0-1:0-45 display TCD and SD behavior, XNFS:0-1:0-0 and XNFS:0-1:0-30 exhibit neo-isotropic responses. All Bi-XNLSs, except for XNFS:0-1:0-15, exhibit near-spherical 3DYMSs. From Fig. 9d, the $\bar{E}$ ranges between 0.014 and 0.018. Notably, XNFS:0-1:0-45 displays the highest $\bar{E}$, while XNFS:0-1:0-30 exhibits the lowest. The E of these Bi-XNLSs closely aligns, falling within the range of 2.75 to 3.50GPa, akin to the stiffness of bone, potentially mitigating the stress shielding effect[45].

Furthermore, we explored the influence of $\bar{\rho}$ at 10%, 20%, and 30%, as elaborated in Supporting Information S3. From Fig. S3, it's evident that $\bar{\rho}$ significantly affects $\bar{S}$ and E compared to its impact on anisotropic behavior[38], indicating that the anisotropic behavior of XNLSs is primarily influenced by the unit cell architectures rather than variations in $\bar{\rho}$[44].

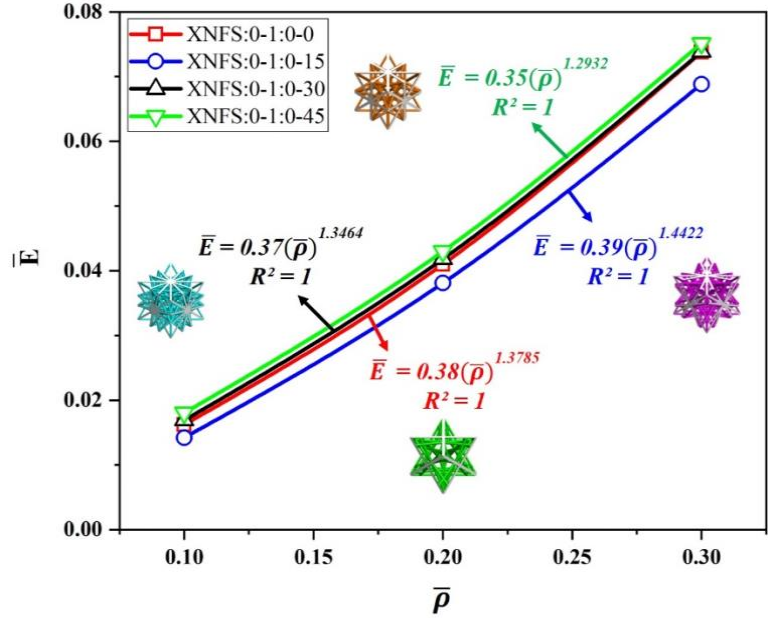

Fig. 10. The performance of Bi-XNLSs: $\bar{E}$ vs $\bar{\rho}$

According to Gibson and Ashby's[9,59,60] scaling law presented in Eq. (9), the architecture-dependent constants $c$ and $n$ govern the deformation behavior of lattice structures.

$$\bar{E} = c\,(\bar{\rho})^n \qquad (9)$$

The lattice structures with $n \approx 1$ and $c \approx 1/3$ primarily exhibit stretching-dominated deformation behavior, enabling them to withstand higher loads and possess high specific stiffness and strength. Conversely, lattices dominated by bending deformation ($n \approx 2$ and $c \approx 1$) exhibit lower load-bearing capacity, along with low specific stiffness and strength. As observed in Fig. 10, Bi-XNLSs have values of $c$ and $n$ closer 1 and 1/3 respectively, indicating that they can be classified as stretching-dominated lattices. This suggests that Bi-XNLSs could prove advantageous for load-bearing bone implants and other structural applications.

3.2.1.2. Tuning anisotropy of Bi-XNLSs

To tune the anisotropic behavior of Bi-XNLSs and pinpoint the instances of perfectly isotropy among them, we conducted a comprehensive parametric study. We varied the strut diameter mismatch $d_0/d_1$ from 0.50 to 1.50 in interval of 0.25, with $d_1$ ranging from 0.6 to 1.0mm with step of 0.1mm, resulting in the design and analysis of 100 lattices. The subsequent section delves deeply into the analysis of XNFS:0-1:0-0. Additional discussions on architectures of XNFS:0-1:0-15, XNFS:0-1:0-30, and XNFS:0-1:0-45 can be found in Support Information S4.

In Fig. 11a, b, each row and column correspond to different $d_0/d_1$ ratios and $d_1$ values, respectively. Each cell presents various attributes (such as unit cell architectures, $\bar{\rho}$, $\bar{S}$, 3DYMSs, Z and $\bar{E}$). The relationship trends among them are plotted in Fig. 11c, d, where solid and dotted lines denote $\bar{\rho}$, Z, $\bar{S}$ and $\bar{E}$, with colors representing different $d_1$ choices.

From Fig. 11a, as the $d_0/d_1$ ratio and $d_1$ increase, strut sizes enlarge, and the available space within the architecture diminishes. At the lowest $d_0/d_1$ ratio and $d_1$, the architecture features smaller strut sizes, resulting in the lowest $\bar{\rho}$ and highest $\bar{S}$. Fig. 11c illustrates that both $\bar{\rho}$ and $\bar{S}$ are more responsive to variations in $d_1$ than $d_0/d_1$ ratios.

Similar to $\bar{\rho}$ and $\bar{S}$, $\bar{E}$ is also influenced by variations in $d_1$ rather than the $d_0/d_1$ ratio[38,39], as depicted in Fig. 11d. As the $d_0/d_1$ ratio increased from 0.50 to 1.00, $\bar{E}$ remained constant, but beyond a ratio of 1.00, it increased. At the lowest $d_0/d_1$ ratio and $d_1$, Fig. 11b shows anisotropic TCD behavior, primarily due to the dominance of $d_1$ over $d_0$. Across different ratios and strut sizes, the anisotropic behavior transitions from TCD to SD. Notably, Z is more sensitive to ratio variations[39] than to strut sizes, as shown in Fig. 11d. With ratios increasing from 0.50 to 1.00, Z increases, then gradually decreases. After a ratio of 1.00, higher strut sizes exhibit a rapid decrease in Z, approaching unity, indicating isotropy. Strut sizes ranging from 0.65 to 0.85 across all ratios show isotropic response. Ultimately, at a $d_0/d_1$ ratio of 0.75 and $d_1$ of 0.7 and 0.8 mm, XNFS:0-1:0-0 demonstrates perfectly isotropic behavior.

This section delved into the performance characteristics of other Bi-XNLSs, where the attributes of XNFS:0-1:0-15, XNFS:0-1:0-30, and XNFS:0-1:0-45 are significantly influenced by $d_1$ compared to $d_0/d_1$ ratios. However, anisotropic behavior is largely affected by $d_0/d_1$ ratios. Both XNFS:0-1:0-0 and XNFS:0-1:0-15 display a smooth transition from SD to TCD behavior, as well as from TCD to SD. However, XNFS:0-1:0-30 and XNFS:0-1:0-45 do not show any transition across ratios and strut sizes, except for all of XNLSs of XNFS:0-1:0-30 demonstrating perfectly isotropic nature, and the XNFS:0-1:0-45 exhibiting a neo-isotropic response.

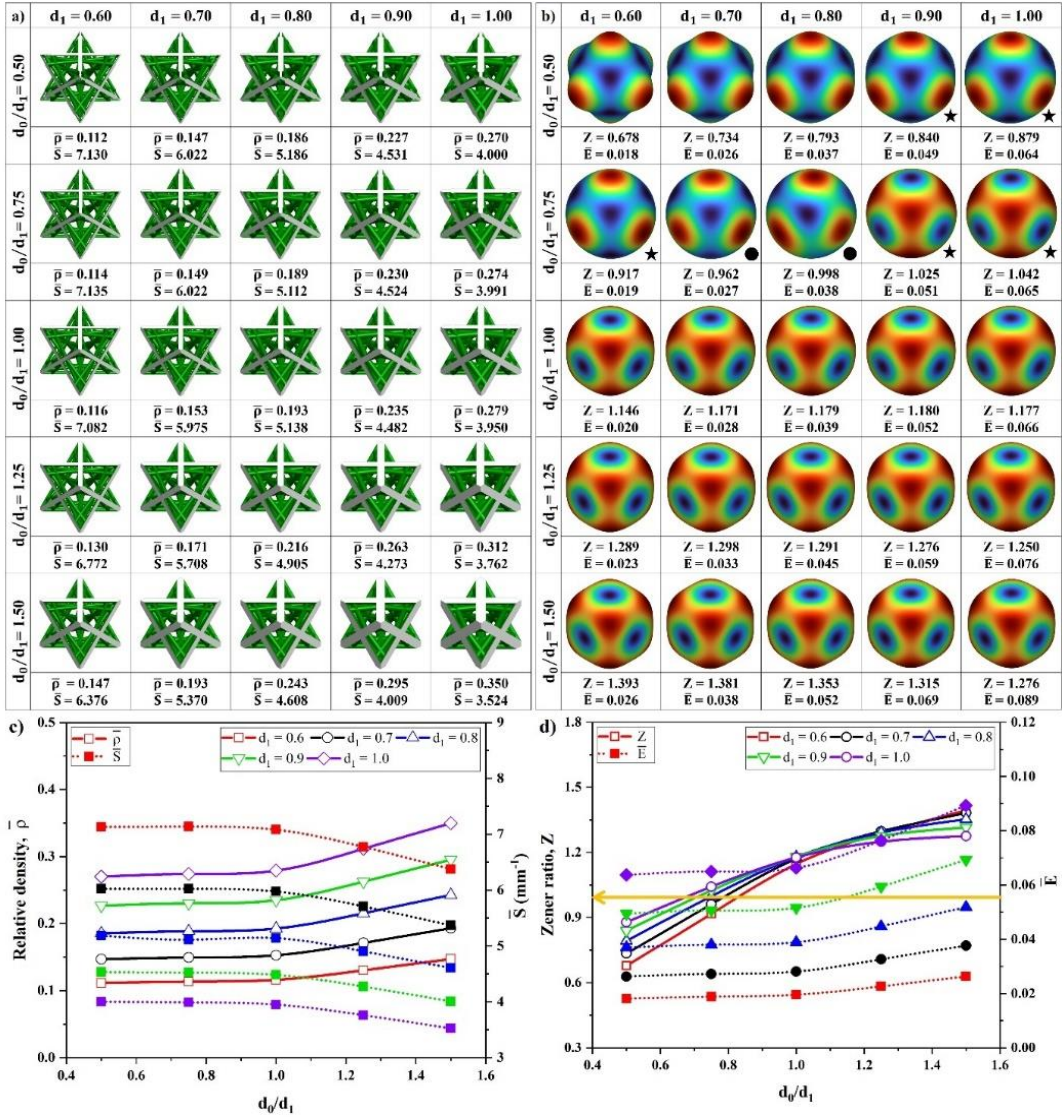

Fig. 11. XNFS:0-1:0-0: a) unit cell architectures, b) 3DYMSs, c) $\bar{\rho}$ and $\bar{S}$, and d) Z and $\bar{E}$

### 3.2.2. Tri-XNLSs

In this section, we delve into Tri-XNLSs, combining all three NOs while changing the $\theta_2$ and $\theta_1$ from 0° to 45° in 15° increments. Sixteen Tri-XNLS configurations are crafted and simulated to analyse their efficacy and showcase their tuneable and controllable anisotropic characteristics. As shown in Fig. 12a, each cell represents 3DYMSs, Z and $\bar{E}$. Each column indicates the $\theta_2$ value and each row represents the XNFS:0-0, XNFS:0-15, XNFS:0-30 and XNFS:0-45 along with $\theta_2$. Tri-XNLSs with $\theta_2$ values of 15° and 30° exhibit elevated $\bar{\rho}$, attributed to an increased number of struts, whereas those with $0^o$ and $45^o$ angles demonstrate lower $\bar{\rho}$, due to fewer struts. Notably, both XNFS:0-15-$\theta_2$ and XNFS:0-30-$\theta_2$ configurations display higher $\bar{\rho}$, compared to others. The collective range of $\bar{\rho}$, across these Tri-XNLSs spans from 0.293 to 0.545.

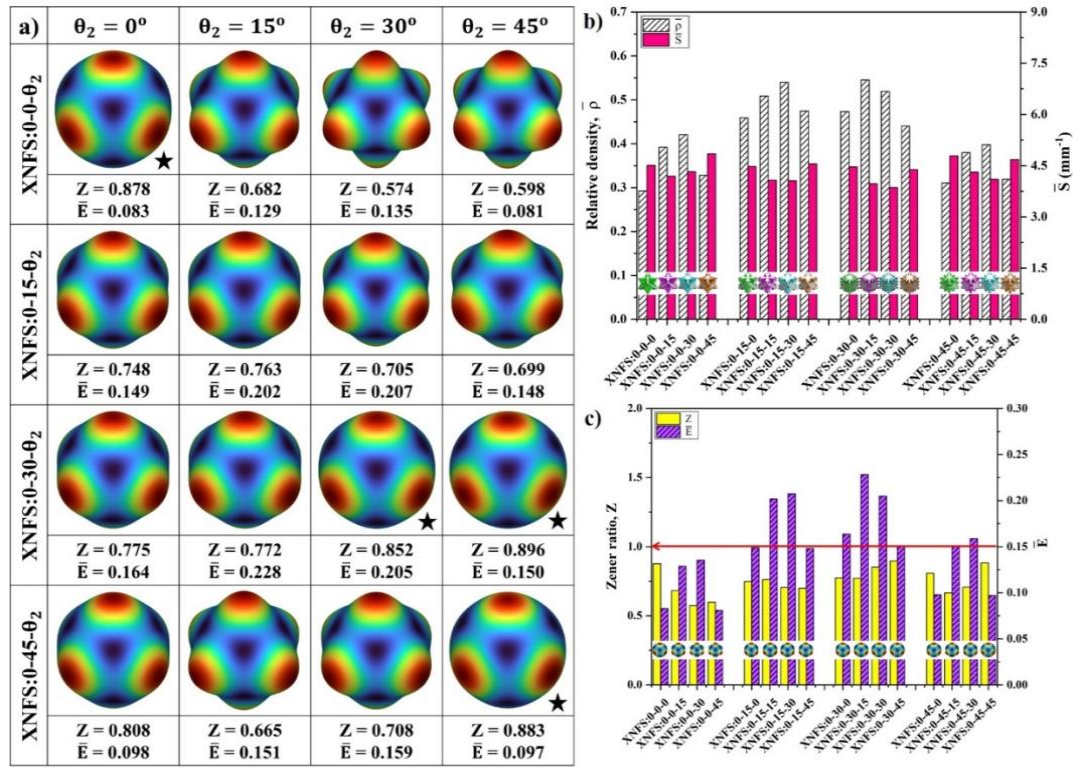

Fig. 12. Performance attributes of Tri-XNLSs: a) 3DYMSs, b) $\bar{\rho}$ and $\bar{S}$, and c) Z and $\bar{E}$

According to Fig. 12b, the $\bar{S}$ of these Tri-XNLSs ranges from 3.859 to 4.848mm$^{-1}$. Notably, at $\theta_2$ values of $0^o$ and $45^o$, all Tri-XNLSs exhibit elevated $\bar{S}$ compared to those at $15^o$ and $30^o$. Interestingly, despite higher $\bar{\rho}$, Tri-XNLSs from both XNFS:0-15-$\theta_2$ and certain XNFS:0-30-$\theta_2$ configurations still maintain $\bar{S}$ above 4mm$^{-1}$ attributed to their increased strut count. This retention of higher $\bar{S}$ proves advantageous for bone implant applications. The combination of both NOs and NORs notably influences $\bar{S}$ and $\bar{\rho}$ (see, Supplementary Information S5).

As depicted in Fig. 12a, it's intriguing to note that Tri-XNLSs continue to demonstrate neo-isotropic behavior, attributable to the amalgamation of various NOs and NORs. The majority of their 3DYMSs appear nearly spherical, although some display TCD behavior due to a higher prevalence of TCD architectures (Supplementary Information S5). The range of Z values across all Tri-XNLSs falls between 0.574 to 0.896. Specifically, Tri-XNLS configurations such as XNFS:0-0-0, XNFS:0-30-30, XNFS:0-30-45, and XNFS:0-45-45 exhibit neo-isotropic behavior, as denoted by a dark star in Fig. 12a. This observation underscores the significant influence of both NOs and NORs on the anisotropic characteristics of Tri-XNLSs. Moreover, it is noteworthy that not only Mono-XNLSs but also Multi-XNLSs showcase isotropic tendencies. In Fig. 12c, it is evident that XNFS:0-30-$\theta_2$ and XNFS:0-15-$\theta_2$ configurations display enhanced $\overline{E}$ in comparison to the rest. With the increase of $\theta_2$ from 0° to 45°, $\overline{E}$ initially rises until reaching 30°, after which it declines, except for XNFS:0-30-$\theta_2$, where $\theta_2$ rises from 0° to 15°, resulting in a rapid increase followed by a decrease for other angles. Across $\theta_2$ values of 15° to 30°, all Tri-XNLSs demonstrate higher $\overline{E}$ relative to others. The overall range of $\overline{E}$ spans from 0.081 to 0.228. Notably, XNFS:0-0-45 and XNFS:0-30-15 exhibit the lowest and highest $\overline{E}$ values, respectively.

### 3.2.2.1. Anisotropic control of Tri-XNLSs

To assess the performance of perfectly isotropic Tri-XNLSs, we employed the method of strut size tuning, as outlined in previous studies[39,44]. This involved conducting a parametric investigation to explore the impact of varying strut sizes. Specifically, these Tri-XNLSs were characterized by three different strut sizes - denoted as $d_0$, $d_1$ and $d_2$ - corresponding to three different NOs. For our analysis, we focused solely on the XNFS:0-0-0 configuration to conduct the comprehensive parametric investigation, which can subsequently be extrapolated to other Tri-XNLSs. This particular configuration was subdivided into three main groups, each maintaining one strut size constant while varying the other two. In the first group, $d_0$ remained constant at 0.8mm, while $d_1$ and $d_2$ were varied ($d_1/d_2$ and varied $d_2$). In the second group, $d_0$ and $d_2$ were varied while keeping $d_1$ constant ($d_0/d_2$ and varied $d_2$). Lastly, in the third group, $d_2$ was kept constant while $d_0$ and $d_1$ were varied ($d_0/d_1$ and varied $d_1$). The ratio of strut sizes in all groups ranged from 0.50 to 1.50, with increments of 0.25, and the actual strut sizes ranged from 0.6 to 1.0mm, with increments of 0.1mm. Each group comprised 25 Tri-XNLSs, resulting in a total of 75 Tri-XNLS configurations that were designed and analysed to evaluate their performance. The performance analysis of the first group is elaborated in the subsequent section, while details regarding the other two groups can be found in Support Information S5.1.

### 3.2.2.1.1. XNFS:0-0-0: Group-1

In Fig. 13a and 13b, each row corresponds to a $d_1/d_2$ ratio, while each column represents the choice of $d_2$. The relationship between these performance metrics and the ratios is illustrated in Fig. 13c and 13d, where solid colored lines denote $\overline{\rho}$ and Z, and colored dotted lines represent $\overline{S}$ and $\overline{E}$, with each color indicating different choices of $d_2$. From Fig. 13c, it's evident that $\overline{\rho}$ is highly responsive to changes in both $d_2$ and the $d_1/d_2$ ratio. The range of $\bar{\rho}$ spans from 0.146 to 0.580. $\bar{\rho}$ increases with an increase in the $d_1/d_2$ ratio, with a gradual increment at lower $d_2$ values and a sharper increase at higher $d_2$ values. This suggests that the $d_1/d_2$ ratio significantly influences $\bar{\rho}$ along with variations in $d_2$. Similarly, $\bar{S}$ is sensitive to changes in both the $d_1/d_2$ ratio and $d_2$, as observed in Fig. 13c. $\bar{S}$ decreases as the $d_1/d_2$ ratio and $d_2$ increase, ranging from 7.433 to 2.084 $mm^{-1}$. This sensitivity is attributed to both $d_1$ and $d_2$ having full cross-sections, thereby impacting $\bar{\rho}$ and $\bar{S}$.

From Fig. 13b, it's noticeable that the 3DYMSs transitioned from a slightly protruded shape along the axes to nearly spherical, indicating a shift from TCD to perfectly isotropic deformation behavior, which aligns with observations from Z. In Fig. 13d, Z demonstrates high sensitivity to changes in the $d_1/d_2$ ratio compared to $d_2$. Furthermore, Fig. 13b illustrates that the neo-isotropic and perfectly isotropic architectures of XNFS:0-0-0 are apparent after reaching a $d_1/d_2$ ratio of 1.00. Specifically, at $d_1/d_2$ ratios of 1.25 and 1.50, most architectures exhibit isotropic behavior across all $d_2$ values. Isotropic architectures are also evident at a $d_1/d_2$ ratio of 1.00 and $d_2$ of 0.6mm, while the remaining configurations demonstrate a neo-isotropic nature.

It is intriguing to note the occurrence of multiple instances of isotropic architectures within XNFS:0-0-0 as the ratios and strut sizes are varied. This observation suggests that introducing strut size mismatches within bioinspired XNLSs yields numerous isotropic XNLS configurations. Specifically, from a $d_1/d_2$ ratio of 1.00 to 1.50, the 3DYMSs in this group adopt spherical shapes. In Fig. 13d, it's evident that as the ratios increase from 0.50 to 1.50, Z rises until the $d_1/d_2$ ratio reaches 1.25, after which some configurations continue to increase while others stabilize. Configurations with $d_2$ values ranging from 0.6 to 0.8mm stabilize, while those with $d_2$ values of 0.9 to 1.0mm increase gradually. Around a $d_1/d_2$ ratio of 1.30 to 1.50, perfectly isotropic XNLSs of XNFS:0-0-0 can be obtained for all $d_2$ values. Moreover, Fig. 13d demonstrates that $\bar{E}$ is influenced by changes in both $d_2$ and $d_1/d_2$ ratios. It gradually increases for lower $d_2$ values of 0.6 to 0.8mm, and for higher $d_2$ values, it increases rapidly as the $d_1/d_2$ ratio increases from 0.50 to 1.00, where $\bar{E}$ ranges from 0.03 to 0.275. Both the geometric attributes and elastic properties of XNFS:0-0-0 are highly reliant on architectural parameters. Not only Z but other properties are also influenced by these classifications of XNFS:0-0-0. This first group exhibits higher ranges of $\bar{\rho}$, $\bar{S}$ and $\bar{E}$ compared to other groups (see, Supplementary Information, S5.1), along with significant variation in anisotropic behavior.

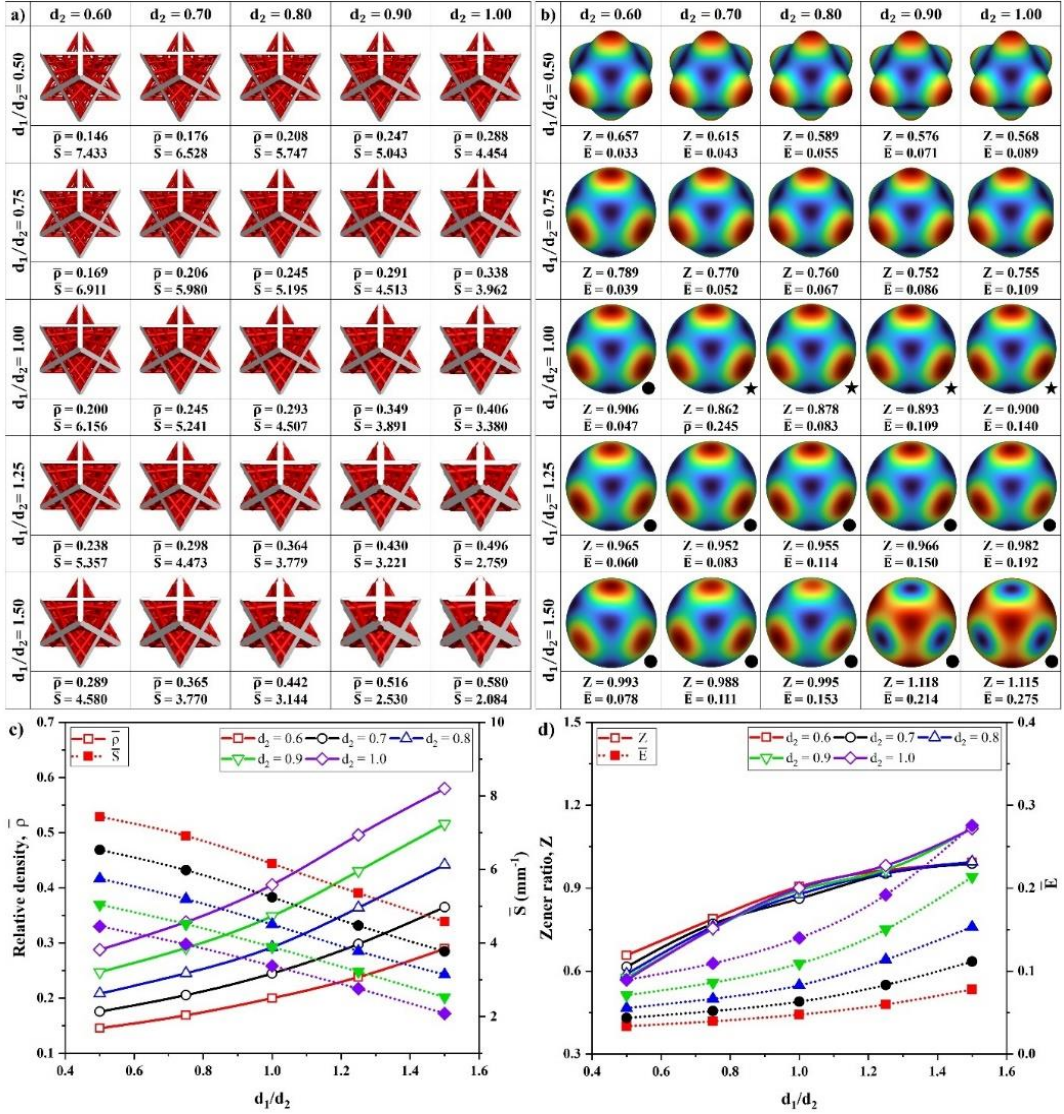

Fig. 13. XNFS:0-0-0 configurations with varying $d_1/d_2$ ratios and $d_2$: a) unit cell architectures, b) 3DYMSs, c) $\bar{\rho}$ and $\bar{S}$, and d) Z and $\bar{E}$

### 3.2.2.1.2. At different $d_0$, $d_1$ and $d_2$

We further subdivided XNFS:0-0-0 into four subgroups based on keeping two strut sizes constant at 0.8mm while varying the third strut size from 0.6 to 1.0mm. This classification allowed for a detailed examination of the parametric study to identify isotropic XNLS configurations and explore their anisotropic behavior and performances. In Fig. 14a and 14b, each row represents subgroups 1 to 4, and each column denotes the strut sizes of the respective subgroups. Fig. 14a illustrates that while the overall shape of XNFS:0-0-0 remains consistent, the bulkiness of struts varies across subgroups. Particularly, subgroup 4 (also called d-subgroup) exhibits uniformly increased bulkiness as the strut size increases.

Fig. 14c showcases the sensitivity of $\bar{\rho}$ to changes in strut size. Across subgroups, $\bar{\rho}$ increases with larger strut sizes, with d-subgroup showing the highest $\bar{\rho}$ ranging from 0.182 to 0.410. Subsequent subgroups exhibit decreasing ranges of $\bar{\rho}$, with the order being subgroup 2 ($d_1$-subgroup), subgroup 1 ($d_2$-subgroup), and subgroup 3 ($d_0$-subgroup). Interestingly, below 0.8mm, the $d_0$-subgroup exhibits the highest $\bar{\rho}$, while the d-subgroup shows the lowest, but this trend reverses after 0.8mm. Similarly, Fig. 14c demonstrates that $\bar{S}$ decreases with increasing strut sizes. The d-subgroup displays the highest range of $\bar{S}$, followed by subgroups of $d_1$, $d_2$ and $d_0$. Remarkably, before 0.8mm, the d-subgroup shows the highest $\bar{S}$, while the $d_0$-subgroup has the lowest, but this trend reverses after 0.8mm.

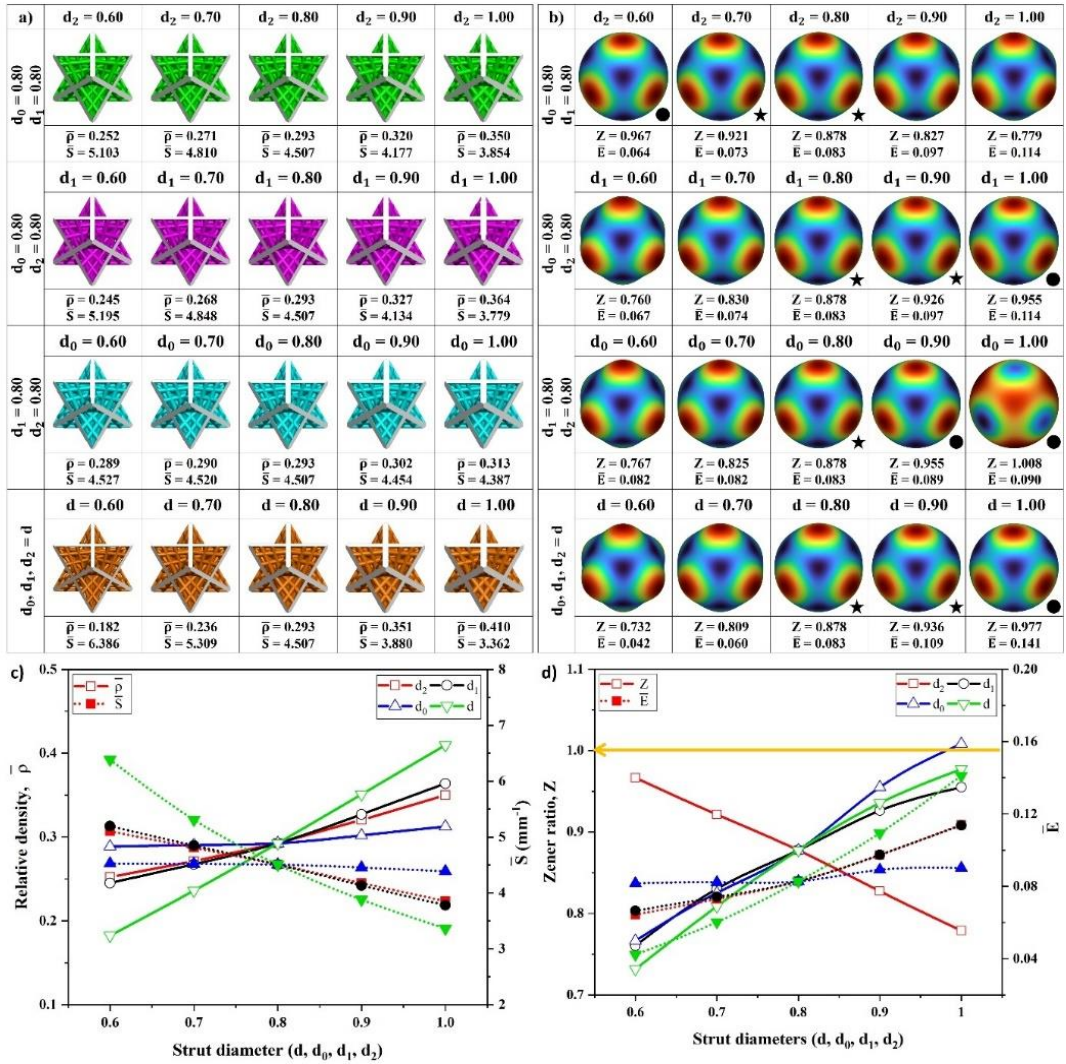

Fig. 14. The properties of subgroups of XNFS:0-0-0: a) unit cell architectures, b) 3DYMSs, c) $\bar{\rho}$ and $\bar{S}$, and d) Z and $\bar{E}$

In Fig. 14b, the 3DYMSs of the $d_2$-subgroup transition from spherical shapes to slightly protruded shapes along axes, while the opposite occurs for other subgroups. The $d_2$-subgroup transitions from perfectly to neo-isotropic and ultimately to TCD behavior as $d_2$ increases. Conversely, the anisotropic behavior of the remaining subgroups transitions from TCD to perfectly isotropic behavior, with Z increasing as their respective strut sizes increase. Fig. 14d illustrates that before 0.8mm, the Z of the $d_2$ subgroup decreases, while the Z of the other subgroups increases. After 0.8mm, the Z of the $d_2$ subgroup further decreases, opposite to the trend observed in other subgroups. $\overline{E}$ follows a trend similar to $\overline{\rho}$, where before 0.8mm, the $d_0$-subgroup has the highest $\overline{E}$, while the d-subgroup has the lowest, but this trend reverses after 0.8mm. Notably, $\overline{E}$ remains unchanged for the $d_2$- and $d_1$-subgroups. Overall, both geometric attributes and elastic properties are influenced by changes in strut sizes across the respective subgroups.

### 3.3. Tuneable and controllable anisotropic behavior of XNLSs

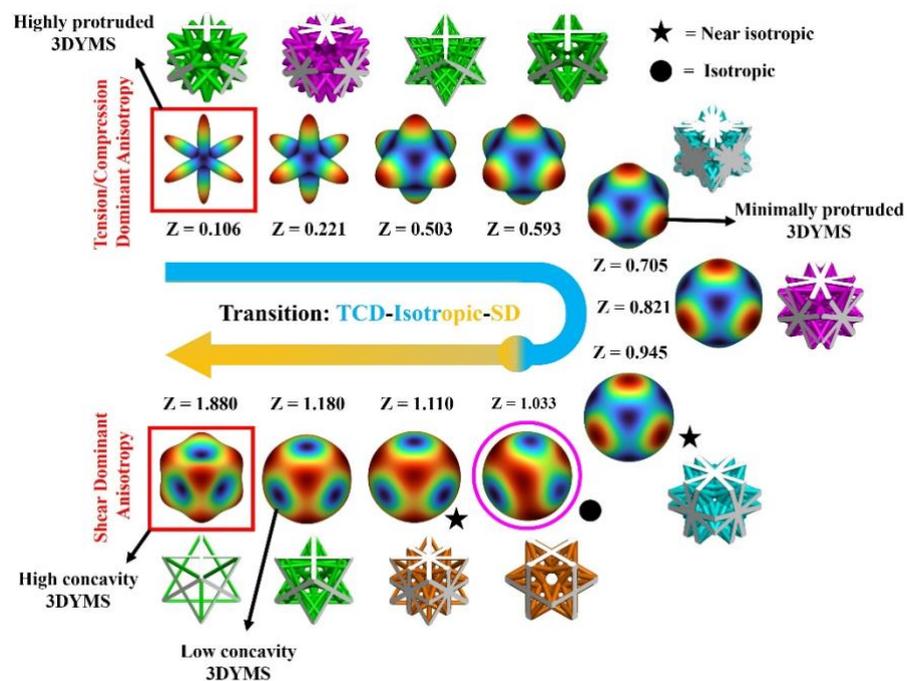

Fig. 15. The transition of anisotropic behavior from Tension/Compression Dominance (blue portion of arrow) to Isotropic to Shear Dominance (orange portion of arrow).

The anisotropic behavior of XNLSs is predominantly dictated by the unit cell architecture[39,44]. XNLSs are categorized into TCD-XNLSs (Z < 1), isotropic-XNLSs (Z = 1), and SD-XNLSs (Z > 1), as illustrated in Fig. 15 where the Z values for all XNLSs are depicted. These classifications entail distinct 3DYMSs, which are independent of the parent material and are delineated in Fig. 15. In instances where the 3DYMSs assume a spherical shape for a given cubic symmetry structure, the XNLSs fall into the category of isotropic-XNLSs, indicated by pink circle and dark dot[39,44]. Conversely, if the 3DYMSs exhibit low to high protrusions or bulges along axes, they belong to the TCD-XNLSs category[44]. Here, an increase in TCD behavior implies heightened anisotropic behavior, indicated by a red square as Z values deviate from unity towards zero. As protrusions evolve into needle-like shapes, XNLSs demonstrate enhanced stiffness along axes, denoted by blue portion of arrow in Fig. 15. Furthermore, if the 3DYMSs showcase low and high depressions or concavities on the six faces of a cube, they are classified as SD-XNLSs, characterized by weak mechanical properties along axes[44]. An increase in shear dominance leads to a departure of Z from unity towards higher values, intensifying the concavity, as illustrated by the orange portion of arrow in Fig. 15. The corresponding XNLS for each 3DYMS is also depicted. Neo-isotropic XNLSs can emerge from both sides of TCD and SD, denoted by dark stars in Fig. 15. This transition serves as a valuable tool for designers in discerning the anisotropic behavior of cubic lattice structures.

In Fig.16a and 16b, our innovative bioinspired lattice designs are contrasted with existing lattice structures[62] at varying relative densities, focusing on normalized elastic modulus and Zener ratio. The Mono-, Bi-, and Tri-XNLSs occupy the red, green, and blue bubbles respectively. The XNLSs span broader ranges of $\bar{\rho}$: Mono-XNLSs range approximately from 0.05 to 0.35, Bi-XNLSs from 0.18 to 0.37, and Tri-XNLSs from 0.28 to 0.55. Regarding $\bar{E}$ shown in Fig. 16a, Mono-, Bi-, and Tri-XNLSs exhibit values within 0.10 to 0.32, 0.20 to 0.30, and 0.25 to 0.42 respectively. Notably, Bi-XNLSs fall between Mono- and Tri-XNLSs, with Mono-XNLSs displaying the widest range of $\bar{E}$. This investigation underscores the broader $\bar{E}$ range (0.10 to 0.42) across $\bar{\rho}$ from 0.05 to 0.55. The majority of these XNLSs cluster closer to neo-isotropic and perfectly isotropic regions, as depicted by the yellow rectangular area in Fig. 16b, with all Bi-XNLSs falling within this region. Some Mono- and Tri-XNLSs also occupy this area. While most Mono- and Tri-XNLSs lie below this region due to exhibiting TCD deformation behavior, they are close to it with a minimum Z value of 0.6. Notably, Mono-XNLSs cover a broader range of Z compared to Tri- and Bi-XNLSs. Given their neo-isotropic and perfectly isotropic characteristics, these bioinspired XNLSs hold promise for numerous applications.

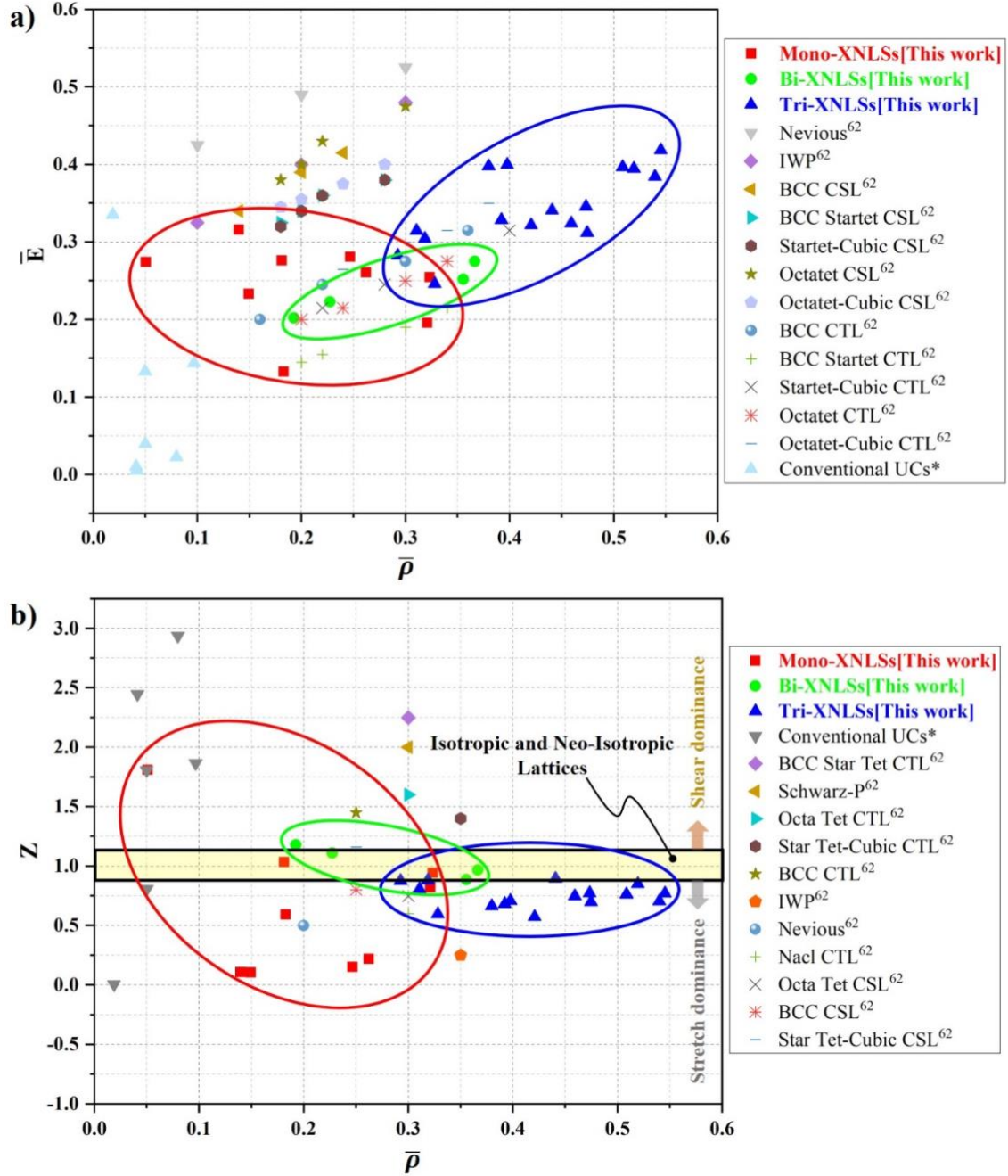

Fig. 16. Comparison of bioinspired XNLSs with extant lattice structures at different relative densities: a) normalized elastic modulus and b) Zener ratio. [*conventional unit cells are designed using same specifications as XNLSs]

## 4. Conclusion

This study presents an innovative lattice design approach inspired by natural bio-architectures observed in cortical bone osteons, golden spirals, and fractals. By integrating architectural elements like "nesting orders" and corresponding "nesting orientations" with the "repetition of self-similar X-cross struts" and "three four-fold axes of symmetry," a wide range of cubic lattices with tuneable and controllable anisotropy are developed. Two new categories of nesting architectures are proposed: Mono-nest lattices, featuring a single nesting order but considering various orientation angles, and Multi-nest (Bi- and Tri-nest) lattices, formed by combining two or three nesting orders with diverse orientation angles for each nested structure.

A total of nine Mono-nest and twenty Multi-nest lattice designs, along with 252 parametric variations, are created and evaluated for their geometric attributes ($\bar{\rho}$ and $\bar{S}$) and elastic properties (Z and $\bar{E}$) through finite element analysis-based numerical homogenization. A comprehensive parametric investigation is conducted to explore the tunability and controllability of anisotropic behavior in Mono-, Bi-, and Tri-nest lattices by varying strut size and strut size mismatch between nested structures. The key findings include:

- Increasing the nesting order of Mono-XNLS from $N_0$ to $N_2$ results in a transition from shear dominant (SD) to tensile/compression dominant (TCD) behavior. Mono-XNLSs derived from $N_1$ at different nesting orientation angles of $\theta_1$ show a smooth transition from TCD to perfectly isotropic behavior, while Mono-XNLSs obtained from $N_2$ and $\theta_2$ exhibit only TCD behavior. The "XNFS:1:45" architecture exhibits perfect isotropy. The strut size variations minimally affect the anisotropic behavior of Mono-XNLSs, but significantly impact $\bar{\rho}$, $\bar{S}$, and $\bar{E}$ properties.

- Most Bi-XNLSs exhibit neo-isotropic and perfectly isotropic behavior, with "XNFS:0-1:0-30" being perfectly isotropic, and other architectures "XNFS:0-1:0-0" and "XNFS:0-1:0-45" being neo-isotropic. Strut size mismatch between nesting orders of Bi-XNLSs have a greater impact on Z than on $\bar{\rho}$, $\bar{S}$, and $\bar{E}$ compared with strut sizes. The anisotropic behavior of Bi-XNLSs can be tuned to achieve perfectly isotropic lattices.

- Tri-XNLSs exhibit neo-isotropic responses, while increasing nesting orders and NOR angles in Multi-XNLSs result in TCD behavior. The "XNFS:0-0-0" architecture exhibits a smooth transition from TCD-isotropic-SD behavior, with its anisotropic behavior highly influenced by strut ratios and variations in strut sizes. Multiple isotropic "XNFS:0-0-0" designs can be generated using optimal parameters for strut sizes.

- Nesting orientations such as 0° and 45° exhibit low $\bar{\rho}$ and $\bar{E}$, but higher $\bar{S}$, demonstrating improved isotropic properties and offering tunability and controllability. Aligning nested lattices with these orientations could benefit applications like bone implants by reducing stress shielding effects, enhancing load-bearing capabilities, and promoting osseointegration.

Traditional lattice structures often lack isotropic behavior, limiting their suitability for various applications. In contrast, the proposed bioinspired nested lattice structures demonstrate tunable and controllable anisotropic behavior across a broad design space, with superior geometric and elastic mechanical properties. Integrating complementary anisotropic behavior through nesting orders and orientations can yield more robust isotropic nested lattices. Future research should explore plasticity, damage behavior, additive manufacturability, and diverse mechanical and functional performances to fully understand these designs. Ultimately, incorporating bioinspired nested lattices into real-world applications such as biomedical, aerospace, and automotive sectors should be pursued due to their exceptional performance and isotropic behavior, particularly in bone implants.

## Acknowledgements

RB would like to acknowledge the MHRD Research Assistantship. SK would like to acknowledge partial financial support from the University of Glasgow through the Reinvigorating Research Award [No: 201644-05]. BP would like to thank the Science and Engineering Research Board (SERB), India for the start-up grant [award no: SRG/2021/000052].

## Competing Interests

Authors declare no financial and non-financial competing interests in the subject matter or materials discussed in this article.

# Supplementary Information

# Bio-inspired nested-isotropic lattices with tuneable and controllable anisotropy

B. Ramalingaiah[1], B. Panda[1], S. Kumar[2,3*]

[1] Department of Mechanical Engineering; Indian Institute of Technology Guwahati, Guwahati-781039, Assam, India,

[2] James Watt School of Engineering, University of Glasgow, Glasgow, G12 8QQ, UK,

[3] Glasgow Computational Engineering Centre, University of Glasgow, Glasgow, G12 8LT, UK

*Email: msv.kumar@glasgow.ac.uk

## S1. Finite element analysis

In the numerical implementation, to evaluate the elements of elasticity tensor, a set linear elastic analyses were f]performed. In each of the analyses, one strain component was assigned a unit value while the remaining five components were set to zero, e.g., see Eq. (S1).

$$\text{input} = \begin{Bmatrix} \varepsilon_{11} \\ \varepsilon_{22} \\ \varepsilon_{33} \\ \varepsilon_{23} \\ \varepsilon_{31} \\ \varepsilon_{12} \end{Bmatrix} = \begin{Bmatrix} 1 \\ 0 \\ 0 \\ 0 \\ 0 \\ 0 \end{Bmatrix}; \ \text{output} = \begin{Bmatrix} \sigma_{11} \\ \sigma_{22} \\ \sigma_{33} \\ \sigma_{23} \\ \sigma_{31} \\ \sigma_{12} \end{Bmatrix} = \begin{Bmatrix} C_{11} \\ C_{21} \\ C_{31} \\ C_{41} \\ C_{51} \\ C_{61} \end{Bmatrix} \qquad (S1)$$

In this methodology, the unit strain was defined as a specified displacement along the boundary, enabling the derivation of corresponding stresses from reaction forces. Consequently, six finite element analyses were conducted to evaluate the components of the stiffness tensor. The boundary conditions were defined for the normal strain $\varepsilon_x$ ($\varepsilon_{11}$) as as follows:

$$\Delta l_{x|x=l_x} = 0.001 l_x$$

$$\Delta l_{x|x=0} = \Delta l_{y|y=l_y} = \Delta l_{y|y=0} = \Delta l_{z|z=l_z} = \Delta l_{z|z=0} = 0 \qquad (S2)$$

which means the displacement in x-axis is $0.001l_x$ when x = $l_x$, i.e., $\varepsilon_x = 0.001$ and the displacement in all other directions is zero.

The boundary conditions for cases involving shear strain were specified differently. For instance, in the scenario of shear strain $\gamma_{xy}$ (=$2\varepsilon_{12}$), the boundary conditions were defined as follows:

$$\Delta l_{x|z=l_z} = 0.0005l_z, \Delta l_{z|x=l_z} = 0.0005l_x$$

$$\Delta l_{z|x=0} = \Delta l_{y|y=l_y} = \Delta l_{y|y=0} = \Delta l_{z|z=l_z} = \Delta l_{x|z=0} = 0$$

(S3)

While the boundary conditions were solely established from a mathematical perspective, this methodology yields exceptionally precise results in predicting the macroscopic mechanical properties of lattice structures.

The volume average of total stress can estimate this stiffness tensor which is given as

$$C_{ij} = \bar{\sigma} = \frac{1}{V} \iiint \sigma_{ij} (x, y, z) \, dV$$

(S4)

The entire homogenization process was executed through finite element analysis, with each numerical test conducted using nTopology.

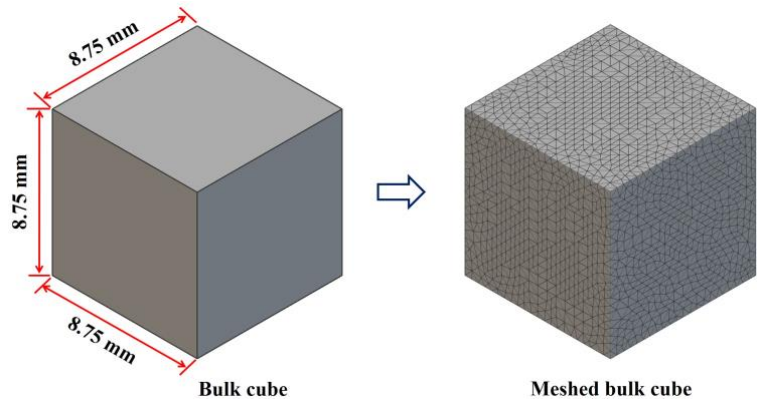

**Fig. S1.1.** Meshed solid cube

In order to validate the numerical homogenization scheme, we employed a solid cube with a unit cell dimension of 8.75×8.75×8.75mm, and meshed it with quadratic tetrahedral elements with an average mesh size of 0.4mm. Fig. S1.1 displays a meshed cube as well as a solid cube. The nTopology software's was utilised to perform numerical homogenization. The elastic properties of 316L stainless steel, including its elastic modulus of 193 GPa and Poisson's ratio of 0.28, were employed for the parent material. In order to determine the E and Z, the fourth order stiffness tensor is extracted and three independent elastic constants are obtained (see Table S1).

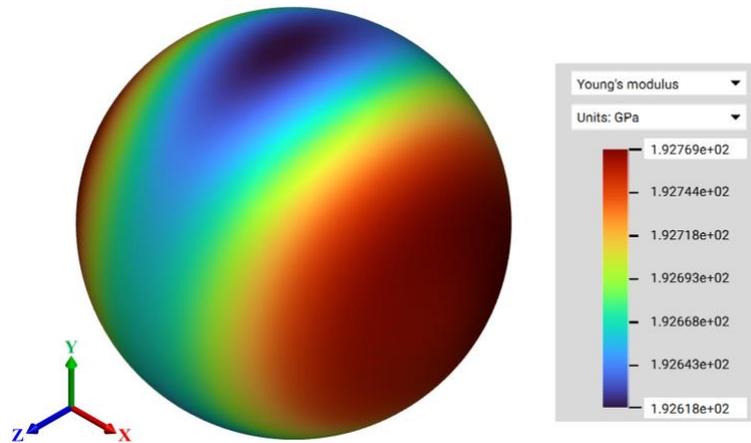

**Fig. S1.2.** 3D Young's modulus surface of bulk 316L SS

**Table S1.** Measured effective elastic properties of bulk 316L SS

| Parameter | $C_{11}$ | $C_{12}$ | $C_{44}$ | $E_s$ | Z |
|-----------|----------|----------|----------|-------|---|
| Unit | GPa | GPa | GPa | GPa | - |
| Value | 246 | 95.5 | 75.3 | 193 | 1 |

From the 3D Young's modulus surface (3DYMS) displayed in Fig. S1.2, it can be seen that the shape of the modulus surface is perfectly spherical and Young's modulus of the solid materials is found to be 193GPa in all directions, giving a Zener number, Z = 1 (see, Table S1), further suggesting that the model employed in this study is valid and can be used to evaluate elastic continuum properties of the lattice structures.

## S2. Mono-XNLSs: Effect of strut sizes

We studied the influence of strut diameter of different nesting orders ($d_0$, $d_1$, and $d_2$) on both geometrical and elastic properties. The diameter ranges from 0.6 to 1.0mm with 0.1mm intervals. We designed and evaluated the performance of 45 Mono-XNLSs. Results are shown in Fig. S2.1 and S2.2, where each row represents Mono-XNLSs and each column represents different choices of strut diameters. The findings in Fig. S2.1a, S2.2a, and S2.2b demonstrate that the size of the struts has a notable impact on $\bar{\rho}$ and $\bar{S}$ which are consistent with the literature[1–3]. Across all Mono-XNLSs, an increase in strut diameter from 0.6 to 1.0mm results in higher $\bar{\rho}$ and lower $\bar{S}$. There is minimal effect on $\bar{\rho}$ with XNFS:0:0, while XNFS:1:15 and XNFS:1:30 show the highest $\bar{\rho}$ and XNFS:0:0 has the highest $\bar{S}$. It can be concluded that NOs, NORs, and strut diameters all significantly influence $\bar{\rho}$ and $\bar{S}$.

Fig. S2.1b and S2.2c illustrate that the anisotropic behavior of Mono-XNLSs is minimally affected by changes in strut diameter as reported elsewhere[1–4]. The most intriguing finding is that XNFS:1:30 approaches isotropy (Z in the range of 0.977 to 0.946) as $d_1$ decreases, while XNFS:1:45 becomes almost perfectly isotropic (Z ranges over 1.062 to 1.006). These two XNLSs exhibit near-perfect spherical shapes of 3DYMSs at different strut sizes.

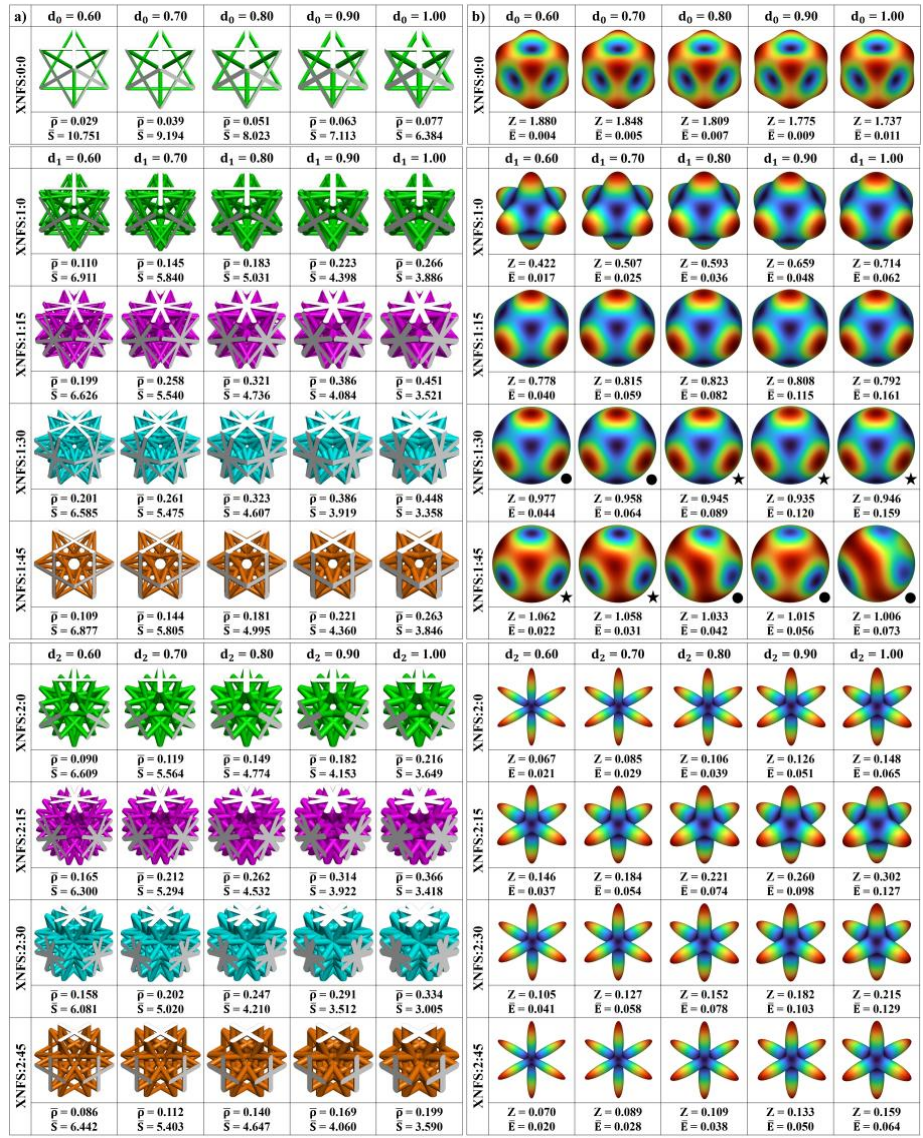

**Fig. S2.1.** Effect of strut sizes on the performance of Mono-XNLSs: a) unit cell architectures and b) 3DYMSs

According to Fig. S2.2d, the $\bar{E}$ value increases as the strut diameter increases, which is consistent with the literature[1–3]. It is worth noting that the highest $\bar{E}$ values are observed for

XNFS:1:15 and XNFS:1:30, XNFS:2:15 and XNFS:2:30 architectures, which are most affected by changes in strut diameter. Conversely, XNFS:0:0 exhibits the lowest $\overline{E}$ value due to having less number of half struts. Finally, XNFS:1:0, XNFS:1:45, XNFS:2:0, and XNFS:2:45 configurations display similar characteristics, with less influence from $d_1$ and $d_2$ diameters.

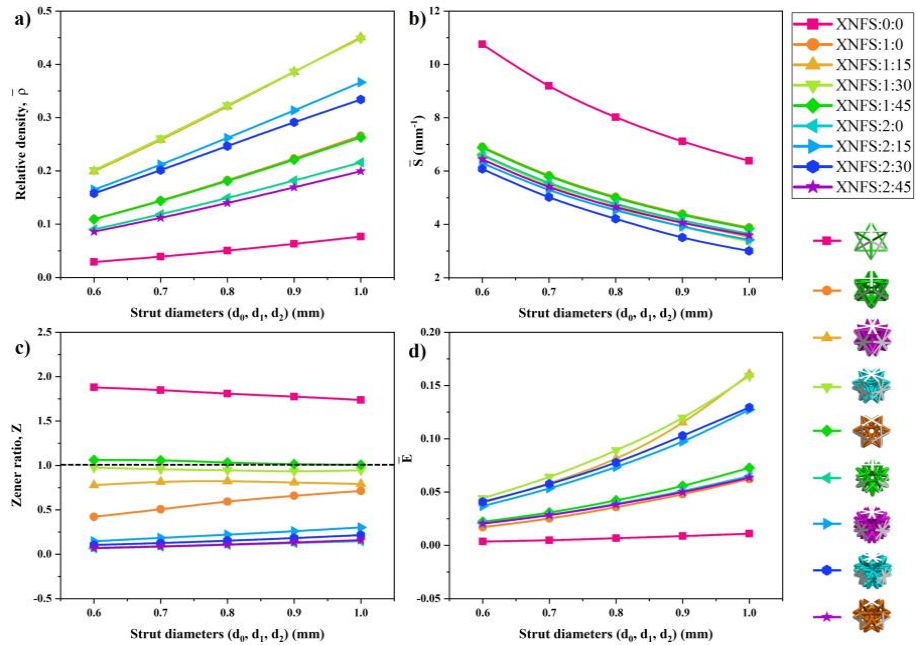

**Fig. S2.2.** Effect of strut diameters ($d_0$, $d_1$, and $d_2$) on the performance attributes of Mono-XNLSs on a) $\overline{\rho}$, b) $\overline{S}$, c) Z and d) $\overline{E}$

## S3. Bi-XNLSs: Effect of relative densities

Herein we examined the influence of different values of $\overline{\rho}$ (0.10, 0.20, and 0.30) on the performance of Bi-XNLSs while maintaining the same strut sizes for both nesting orders. To do so, a total of nine designs were created and analysed in order to assess their properties. The results are presented in Fig. S3.

Fig. S3a and S3c demonstrates that as $\overline{\rho}$ increases, the architectures become bulkier and the space within them decreases due to larger strut sizes, leading to a decrease in $\overline{S}$. Fig. S3c shows that XNFS:0-1:0-15 and XNFS:0-1:0-30 have similar strut diameters and $\overline{S}$, with higher $\overline{S}$ and lower strut size. These findings suggest that $\overline{\rho}$ has a greater impact on $\overline{S}$, which is more affected by strut sizes. When $\overline{\rho}$ is higher (0.30), still all Bi-XNLSs have a higher $\overline{S}$ (4-7mm$^{-1}$), making them more advantageous for bone implant applications.

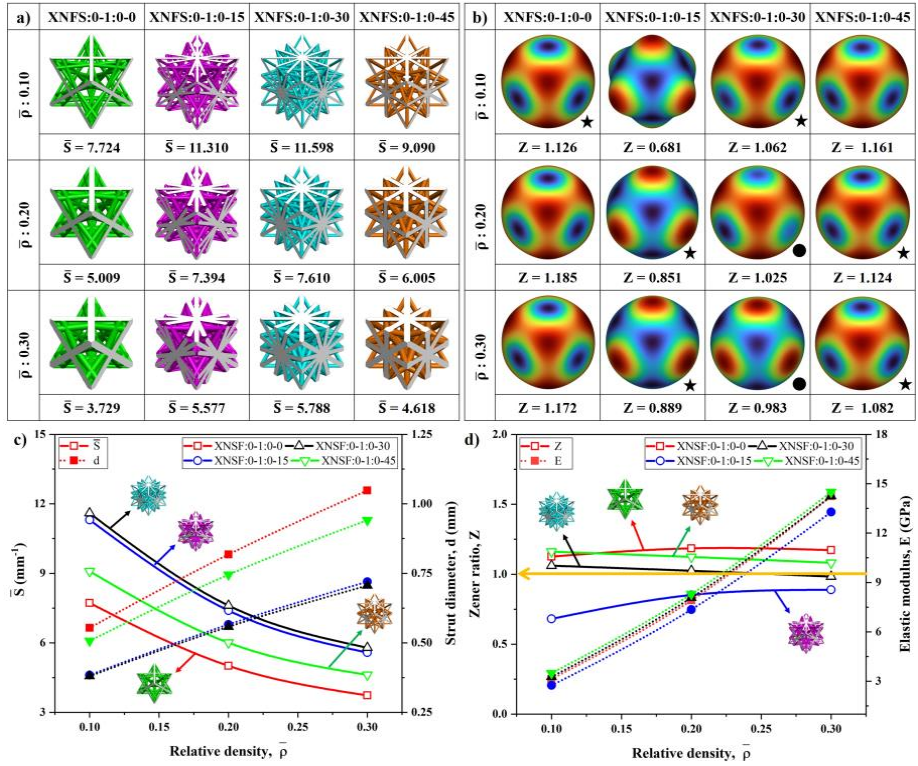

**Fig. S3.** Performances Bi-XNLSs for different $\bar{\rho}$ of 0.10, 0.20 and 0.30: a) unit cell architectures, b) 3DYMSs, c) $\bar{S}$ and strut diameters, and d) Z and E

Based on the observation from Fig. S3b and 3d, it can be concluded that the $\bar{\rho}$ has less impact on the anisotropic behavior of the Bi-XNLSs. The shape of the 3DYMSs only slightly changes as the $\bar{\rho}$ is increased. For XNFS:0-1:0-0, Z increases initially from $\bar{\rho}$ of 0.10 to 0.20 and then decreases from 0.20 to 0.30. On the other hand, all other Bi-XNLSs tend to become more isotropic as $\bar{\rho}$ is increased. XNFS:0-1:0-15 transitions to isotropic nature from the TCD behaviour, whereas XNFS:0-1:0-30 becomes isotropic from the neo-isotropic behavior. For XNFS:0-1:0-45, it becomes neo-isotropic from the SD nature. Therefore, it can be concluded that the anisotropic behavior of Bi-XNLSs moves towards isotropic nature as the $\bar{\rho}$ is increased, either from the TCD or SD behaviour.

Based on previous studies[2,3,5], it is evident from Fig. S3b that $\bar{\rho}$ has a significant influence on E. This effect is demonstrated in Fig. S3d, where increasing $\bar{\rho}$ results in a rapid increase in E. All Bi-XNLSs show similar trends in E, with XNFS:0-1:0-45 having the highest E, followed by XNFS:0-1:0-30, XNFS:0-1:0-0, and XNFS:0-1:0-15. For $\bar{\rho}$ values ranging from 0.10 to 0.30, E falls between 3 GPa and 15 GPa, making it ideal for bone implant applications. Lastly, $\bar{\rho}$ has a greater impact on strut diameters and E, with a lesser effect on anisotropic behavior.

## S4. Bi-XNLSs: Tuning and controlling anisotropic behaviour

### S4.1. XNFS:0-1:0-15 group

In Fig. S4.1a and b, each row displays $d_0/d_1$ ratios ranging from 0.50 to 1.50 with an increment of 0.25, while each column showcases strut sizes of $d_1$ from 0.6 to 1.0mm with an increment of 0.1mm. With an increase in $d_0/d_1$ ratio and strut size, the strut bulkiness also increases, resulting in a decrease in the void volume fraction of the unit cell. When the ratio and strut size are lower, smaller cross sections of struts lead to the lowest value of $\bar{\rho}$ and the highest value of $\bar{S}$. However, the opposite occurs when they are higher.

Fig. S4.1c shows that changes in strut size have a greater impact on both $\bar{\rho}$ and $\bar{S}$ compared to changes in $d_0/d_1$ ratio. As the ratio increases from 0.50 to 1.50, $\bar{\rho}$ gradually increases and $\bar{S}$ decreases. Within the range of these ratio and strut size variations, $\bar{\rho}$ ranges from 0.205 to 0.545, and $\bar{S}$ ranges from 7.027 to 3.154mm$^{-1}$. The $\bar{E}$ is also more affected by changes in $d_1$ compared to changes in the $d_0/d_1$ ratio and ranges from 0.041 to 0.204, as seen in the trends of Fig. S4.1d.

Based on Fig. S4.1b, it can be observed that the anisotropic behavior of XNFS:0-1:0-15 changes from TCD to isotropic, as seen from their 3DYMSs and Z values. Once the $d_0/d_1$ ratio reaches 1.00, the architectures evince neo-isotropic behaviour at different $d_1$ values, where they have near-perfect spherical shapes of 3DYMSs. As XNFS:0:0 and XNFS:1:15 are combined, at lower ratios and strut sizes, the dominance of XNFS:1:15 is high, leading to TCD behaviour. However, as the ratio increases, the dominance of XNFS:0:0 increases, leading to SD response. This study considered the ratio up to 1.50 and $d_1$ of 1.0mm, resulting in isotropic behavior. Based on Fig. S4.1b and S4.1d, it is observed that Z is sensitive to the $d_0/d_1$ variations rather than changes in $d_1$, which is consistent with literature[1]. The isotropic XNFS:0-1:0-15 is found at the $d_0/d_1$ ratio of 1.25 and $d_1$ of 1.0mm, $d_0/d_1$ ratio of 1.50 and $d_1$ of 0.6, 0.7, and 1.0mm, while the near isotropic behavior is found for the ratio of 1.00, 1.25, and 1.50 at different $d_1$ values.

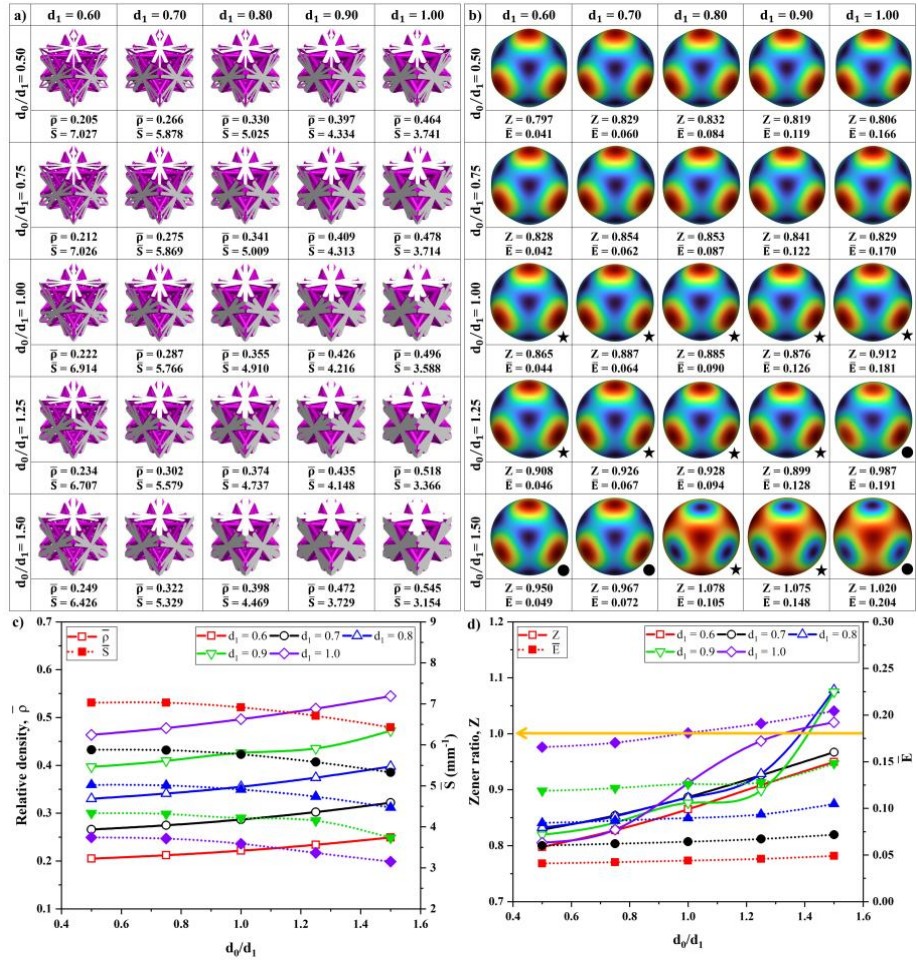

**Fig. S4.1.** XNFS:0-1:0-15 group: a) unit cell architectures, b) 3DYMSs, c) $\bar{\rho}$ and $\bar{S}$, and d) Z and $\bar{E}$

### S4.2. XNFS:0-1:0-30 group

In Fig. S4.2c, it was discovered that both $\bar{\rho}$ and $\bar{S}$ are sensitive to changes in $d_1$, compared to changes in $d_0/d_1$ ratio. As the ratio increased from 0.50 to 1.50, $\bar{\rho}$ gradually increased and $\bar{S}$ gradually decreased. The range of $\bar{\rho}$ is from 0.208 to 0.584, and $\bar{S}$ is in the range of 7.060 to 3.395mm$^{-1}$. In Fig. S4.2d, it was found that $\bar{E}$ is sensitive to changes in $d_1$, compared to changes in the ratio $d_0/d_1$. As the ratio increased from 0.50 to 1.50, $\bar{E}$ gradually increased and was found in the range of 0.045 to 0.215.

One interesting feature of XNFS:0-1:0-30 is that all of its 3DYMSs are spherical in shape, with their Z values being close to 1, as shown by the dark dots in Fig. S4.2b. With a ratio range of 0.50 to 1.50 and a $d_1$ of 0.6 to 1.0mm, all unit cell architectures exhibit isotropic behavior, making them advantageous for bone implants and other applications. According to Fig. S4.2d, the Z values of these architectures are between 0.95 and 1.05, indicating that they are isotropic XNLSs.

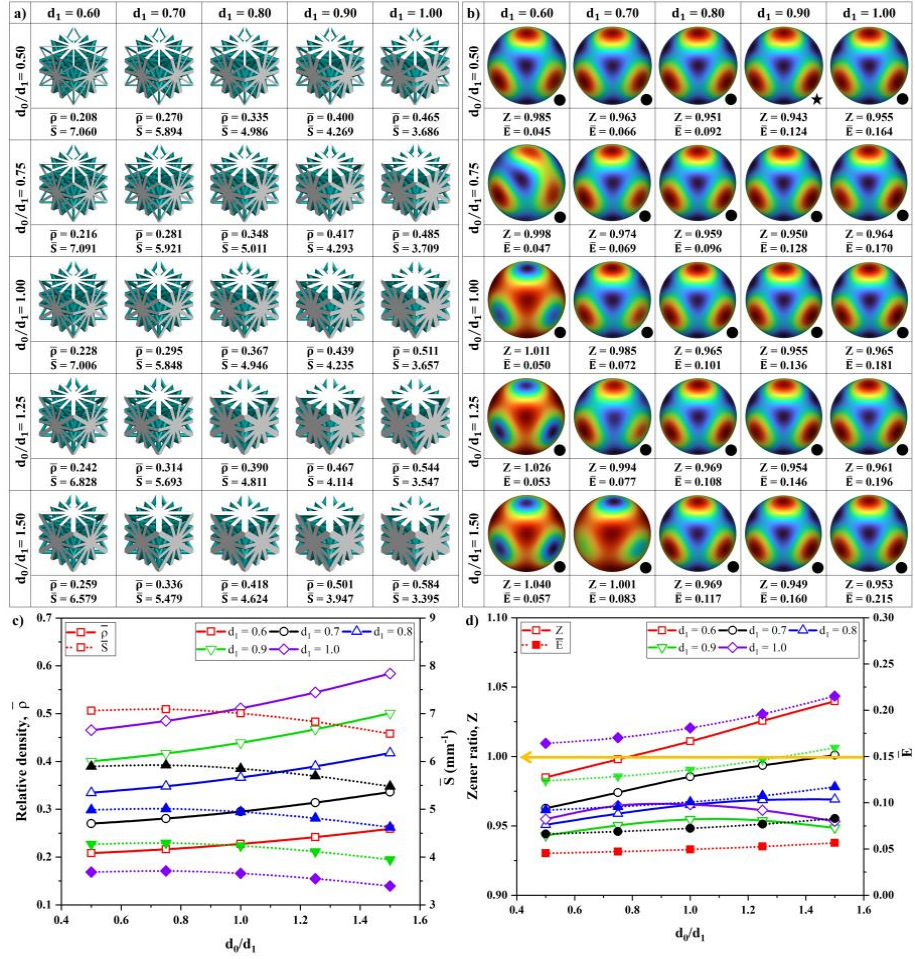

**Fig. S4.2.** XNFS:0-1:0-30 group: a) unit cell architectures, b) 3DYMSs, c) $\bar{\rho}$ and $\bar{S}$, and d) $Z$ and $\bar{E}$

### S4.3. XNFS:0-1:0-45 group

Fig. S4.3c shows that $\bar{\rho}$ and $\bar{S}$ are highly sensitive to changes in $d_1$ compared to changes in $d_0/d_1$ ratio. The range of $\bar{\rho}$ is between 0.116 and 0.410, and the range of $\bar{S}$ is between 7.758 and 3.842 mm$^{-1}$. As $d_0/d_1$ ratio increases from 0.50 to 1.50, $\bar{\rho}$ increases gradually, and $\bar{S}$ decreases. Fig. S4.3d shows $\bar{E}$ ranges from 0.023 to 0.111 and is dependent on changes in $d_1$ compared to changes in $d_0/d_1$ ratio. As $d_0/d_1$ ratio increases from 0.50 to 1.50, $\bar{E}$ increases gradually at a lower rate for the lower $d_1$ values and a slightly higher rate for the higher $d_1$ values.

Based on the findings from Fig. S4.3b, it can be observed that the 3DYMSs of XNFS:0-1:0-45 are mostly of near-perfect and perfect spherical shape. For ratios below 1.00, most of them are neo-isotropic and perfectly isotropic. However, some of them exhibit neo-isotropic behavior after the ratio of 1.00 for higher $d_1$ sizes. From Fig. S4.3d, it was observed that most of the architectures lie between the Z values of 1.0 to 1.1, which can be considered as neo-isotropic and perfectly isotropic behavior. It was also noticed that Z is more sensitive to changes in ratios than changes in $d_1$ sizes. At the ratio of 0.50 and $d_1$ of 0.90 and 1.00mm, and at the ratio of 0.75 and $d_1$ of 1.00mm, XNFS:0-1:0-45 exhibited perfectly isotropic behavior.

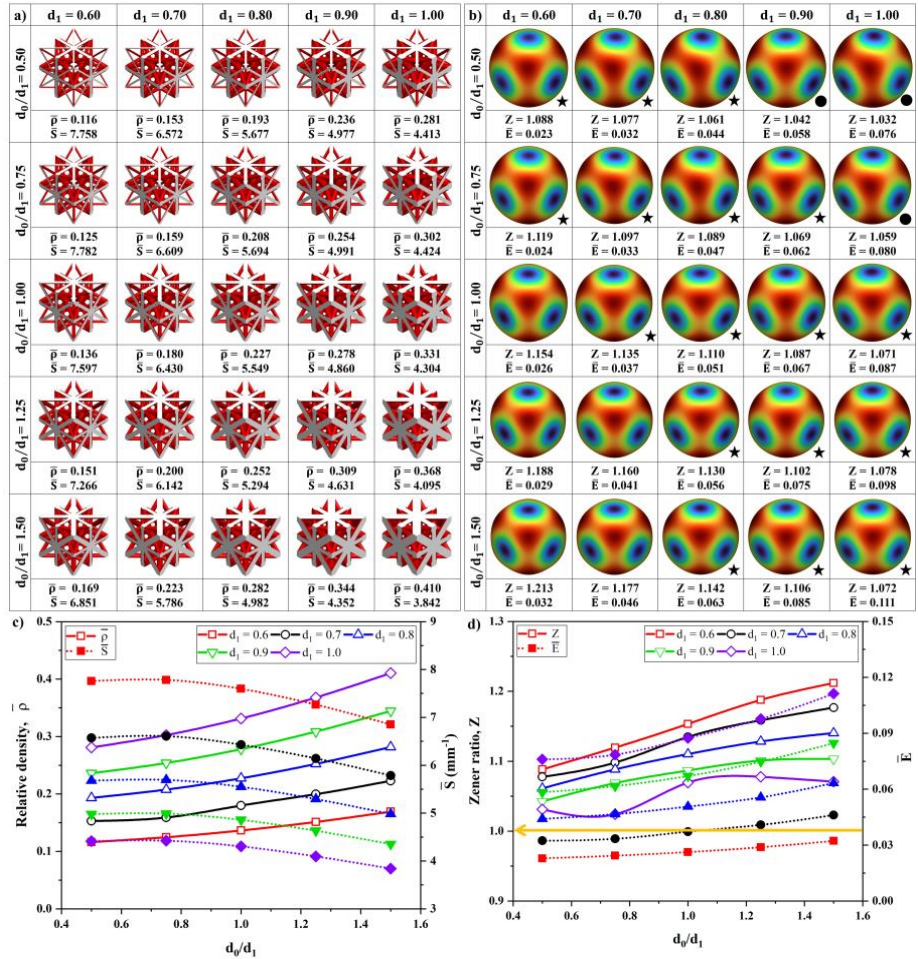

**Fig. S4.3.** XNFS:0-1:0-45 group: a) unit cell architectures, b) 3DYMSs, c) $\bar{\rho}$ and $\bar{S}$, and d) Z and $\bar{E}$

## S5. Tri-XNLSs

Fig. 6c illustrates the variation in unit cell architectures of Tri-XNLSs as $\theta_2$ changes. Tri-nest lattices show an increase in both struts and nodes, enhancing strut connectivity and relative density. However, this also leads to greater heterogeneity in pore sizes. Notably, XNFS:0-15-$\theta_2$ and XNFS:0-30-$\theta_2$ display higher geometric attributes compared to XNFS:0-0-$\theta_2$ and XNFS:0-45-$\theta_2$. Tri-XNLSs at $\theta_2$ between 15° to 30° demonstrate superior performance compared to other $\theta_2$ choices. Observing Tri-XNLSs reveals that both NOs and NORs significantly influence architectural parameters.

Fig. 12b presents $\bar{\rho}$ of Tri-XNLSs. As $\theta_2$ increases from 0° to 15°, $\bar{\rho}$ rises more steeply for all Tri-XNLSs (Tri-XNLSs of XNFS:0-0-$\theta_2$, XNFS:0-15-$\theta_2$, XNFS:0-30-$\theta_2$ and XNFS:0-45-$\theta_2$). Subsequently, it gradually increases from 15° to 30° for all except XNFS:0-30-$\theta_2$, before decreasing more rapidly from 30° to 45°. The $\bar{\rho}$ range for XNFS:0-0-$\theta_2$ varies from 0.293 to 0.328, for XNFS:0-15-$\theta_2$ from 0.459 to 0.540, for XNFS:0-30-$\theta_2$ from 0.441 to 0.545, and finally for XNFS:0-45-$\theta_2$ from 0.310 to 0.398 across different $\theta_2$ values.

For XNFS:0-0-$\theta_2$, as $\theta_2$ increases from 0° to 30°, Z decreases and the deformation behavior transitions from near isotropic to TCD, then increases at 45°. Conversely, for XNFS:0-15-$\theta_2$, there's minimal change in anisotropic behavior, as reflected in slight alterations in 3DYMSs, indicating a transition from TCD to neo-isotropic behavior as $\theta_2$ increases from 0° to 45°. However, XNFS:0-45-$\theta_2$ exhibits a distinct pattern: as $\theta_2$ increases from 0° to 15°, it shifts from neo-isotropic to TCD behavior, maintains TCD from 15° to 30°, and then returns to neo-isotropic from 30° to 45°.

### S5.1. Tri-XNLSs: Tuning and controling of anisotropic behaviour

S5.1.1. XNFS:0-0-0:Group-2

In Fig. S5.1a and b, each row corresponds to the $d_0/d_2$ ratio (ranging from 0.50 to 1.50), while each column represents variations in $d_2$ (ranging from 0.6 to 1.0mm). As depicted in Fig. S5.1c, $\bar{\rho}$ demonstrates high sensitivity to changes in $d_2$, spanning from 0.244 to 0.423. Initially stable, $\bar{\rho}$ begins to increase beyond a ratio of 1.00. Similarly, $\bar{S}$ shows high sensitivity to $d_2$ variations, maintaining stability until a ratio of 1.00 before decreasing, ranging from 5.125 to 5.125-3.460 mm$^{-1}$.

According to Fig. S5.1d, $\bar{E}$ is notably affected by variations in the $d_0/d_2$ ratio. It remains consistent within the range of 0.50 to 1.00 but steadily rises thereafter. With increasing $d_0/d_2$ ratio, $\bar{E}$ increases from 0.061 to 0.146. In Fig. S5.1b, 3DYMSs transition from a low axially protruded shape to a near-spherical one, indicating a smooth shift from TCD to near isotropic behavior. Fig. S5.1d reveals Z's sensitivity to changes in the $d_0/d_2$ ratio, with near isotropic architectures observed at a ratio of 1.00. At a ratio of 1.25, most architectures exhibit isotropic behavior within the range of $d_2$ from 0.7 to 1.0mm. Perfectly isotropic XNLSs of XNFS:0-0-0 can be achieved around a ratio of 1.15 to 1.35 for $d_2$ within the same range.

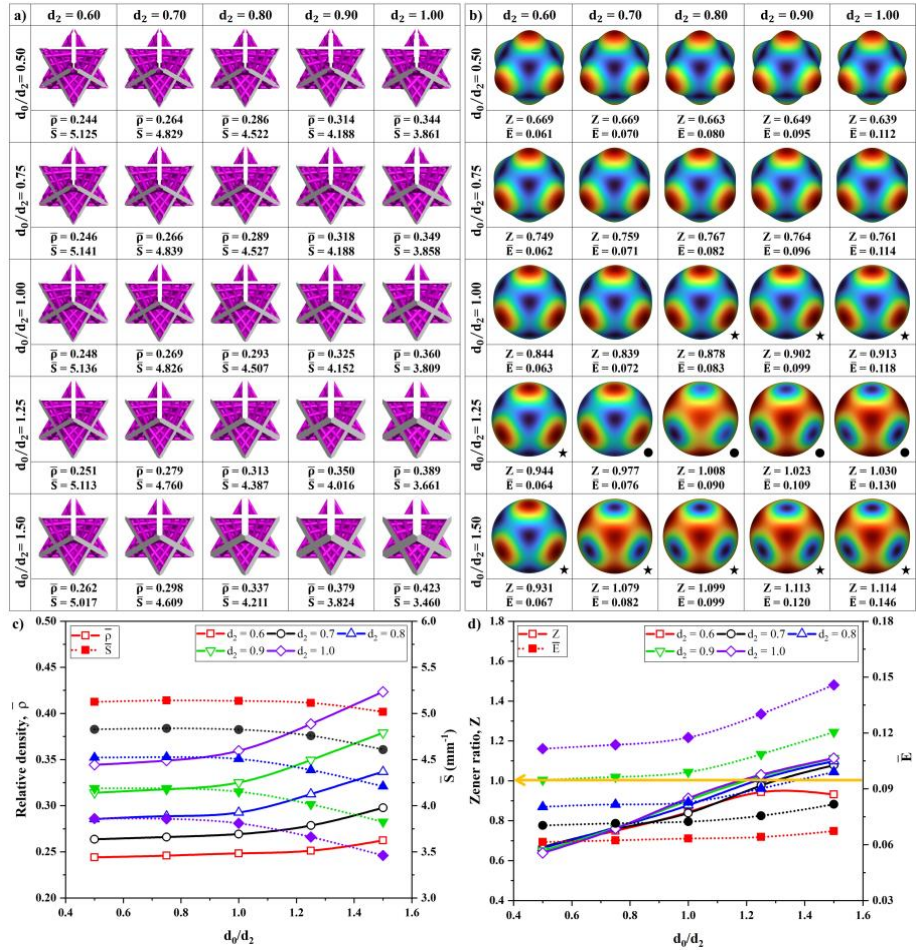

**Fig. S5.1.** The properties of XNFS:0-0-0 at different $d_0/d_2$ ratios and $d_2$ choices: a) unit cell architectures, b) 3DYMSs, c) $\bar{\rho}$ and $\bar{S}$, and d) Z and $\bar{E}$

S5.1.2. XNFS:0-0-0:Group-3

In Fig. S5.2a and b, each row represents the ratio of $d_0/d_1$, and each column represents the variation of $d_1$. Fig. S5.2c shows that $\bar{\rho}$ is highly sensitive to changes in $d_1$ compared to changes in the $d_0/d_1$ ratio. The range of $\bar{\rho}$ is between 0.231 to 0.431. $\bar{\rho}$ remains constant until the $d_0/d_1$ ratio reaches 1.00, and then it increases. The increment is lower for smaller $d_1$ and higher for larger $d_1$ due to the presence of a half strut in $d_0$. Similarly, $\bar{S}$ is also highly sensitive to changes in $d_1$ compared to changes in the $d_0/d_1$ ratio, as observed in Fig. S5.2c. As $d_0/d_1$ ratio and $d_1$ increase, $\bar{S}$ varies from 5.247 to -3.423mm$^{-1}$. Fig. S5.2d shows that changes in the $d_1$ compared with $d_0/d_1$ ratios affect the $\bar{E}$ where it remains stable from $d_0/d_1$ ratios of 0.50 to 1.00, and then gradually increases after ratios of 1.00. As $d_0/d_1$ ratios and $d_1$ increase, $\bar{E}$ changes from 0.061 to 0.143.

According to Fig. S5.2b, the 3DYMss of unit cell architectures change from an axially protruded shape to a near-spherical shape, indicating a smooth transition of anisotropic behavior. Fig. S5.2d reveals that isotropic XNLSs are found at the $d_0/d_1$ ratios of 1.00, 1.25, and 1.50. This is an interesting observation as it shows that by keeping $d_2$ constant and changing the other two strut sizes, multiple isotropic architectures can be obtained due to the combination of struts of three different NOs. This group of architectures also exhibit neo-isotropic behavior at different ratios. As the ratio increases from 0.50 to 1.50, the Z value gradually increases, and for strut sizes of 0.7, 0.8, 0.9, and 1.0mm, the exact isotropic XNLSs for XNFS:0-0-0 can be obtained at a ratio of around 0.90 to 1.50.

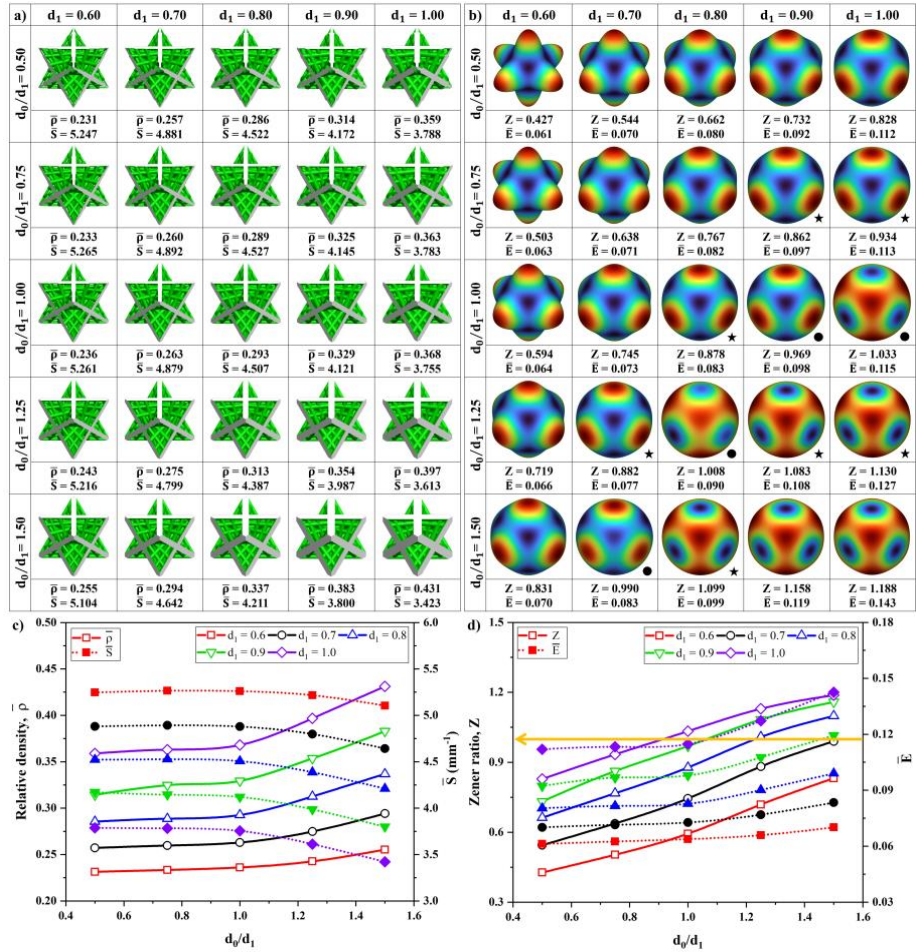

**Fig. S5.2.** The properties of XNFS:0-0-0 at different $d_0/d_1$ ratios and $d_1$ choices: a) unit cell architectures, b) 3DYMSs, c) $\bar{\rho}$ and $\bar{S}$, and d) Z and $\bar{E}$